\pdfoutput=1
\documentclass[11pt,reqno,preprint]{article}
\usepackage{jheppub,epsfig,amssymb,amsmath,mathrsfs,hyperref,graphicx}
\usepackage{multirow,bbm}
\usepackage[makeroom]{cancel}
\usepackage{caption}
\usepackage{subcaption}

\def\be{\begin{equation}}
\def\ee{\end{equation}}
\def\ba{\begin{eqnarray}}
\def\ea{\end{eqnarray}}
\newcommand{\bea}{\begin{eqnarray}}
\newcommand{\eea}{\end{eqnarray}}
\newcommand{\nn}{\nonumber}

\def\Li{\textrm{Li}}

\def\eqn#1{eq.~(\ref{#1})}

\def\eqn#1{eq.~(\ref{#1})}

\newcommand{\fwboxL}[2]{\text{\makebox[#1][l]{$#2$}}}

\def\lr{\leftrightarrow}
\def\RR{\mathcal{R}}
\def\EE{\mathcal{E}}

\def\Gcusp{\Gamma_{\rm cusp}}

\def\tr{{\rm tr}}
\title{A Two-Loop Four-Point Form Factor at Function Level}

\author{Lance~J.~Dixon$^{1}$}
\author{and Shuo Xin$^{1}$}

\affiliation{$^1$ SLAC National Accelerator Laboratory,
Stanford University, Stanford, CA 94309, USA}

\abstract{Recently, the maximally-helicity-violating four-point
form factor for the chiral stress-energy tensor in planar
$\mathcal{N}=4$ super Yang-Mills was computed to three loops
at the level of the symbol associated with multiple polylogarithms.
It exhibits {\it antipodal self-duality}, or
invariance under the combined action of a kinematic map and
reversing the ordering of letters in the symbol.
Here we lift the two-loop form factor from symbol level to function level.
We provide an iterated representation of the function's derivatives
(coproducts).  In order to do so, we find a three-parameter limit of
the five-parameter phase space where the symbol's letters are all rational.
We also use function-level information about dihedral symmetries
and the soft, collinear, and factorization limits,
as well as limits governed by the form-factor operator product expansion
(FFOPE).  We provide plots of the remainder function on several kinematic
slices, and show that the result is compatible with the FFOPE data.
We further verify that antipodal self-duality is valid at two loops
beyond the level of the symbol.}

\emailAdd{lance@slac.stanford.edu}\emailAdd{xinshuo@stanford.edu}

\preprint{ \begin{flushright} SLAC--P--24002 \end{flushright}}

\begin{document}
\hypersetup{pageanchor=false}
\maketitle
\hypersetup{pageanchor=true}

\section{Introduction}
\label{sec:intro}

Perturbative calculations of scattering amplitudes and form factors in
planar $\mathcal{N}=4$ super-Yang-Mills (SYM) theory have seen tremendous
progress in recent years. Among many recent developments, bootstrap methods
have proven to be particularly effective in pushing the results to both
high multiplicity and high loop orders~\cite{%
Dixon:2011pw,Dixon:2011nj,Dixon:2013eka,Dixon:2014voa,Dixon:2014iba,
Drummond:2014ffa,Dixon:2015iva,Caron-Huot:2016owq,Dixon:2016apl,
Dixon:2016nkn,Drummond:2018caf,Caron-Huot:2019vjl,Dixon:2020bbt,
Dixon:2022rse,Dixon:2023kop,Basso:2024hlx}.
A critical ingredient for such methods is sufficient boundary-value data,
which can be supplied by the flux-tube representation or pentagon
operator-product expansion (OPE)~\cite{%
Basso:2013vsa,Basso:2013aha,Basso:2014koa,Basso:2014nra,Basso:2015rta,
Basso:2015uxa}.  This representation has recently been extended to
form factors of protected operators
(FFOPE)~\cite{Sever:2020jjx,Sever:2021nsq,Sever:2021xga,Basso:2023bwv}.
Another source of boundary-value data for amplitudes is multi-Regge
kinematics~\cite{Balitsky:1978ic,Kuraev:1977fs}, which has been understood
to all subleading logarithms for six-point scattering~\cite{Basso:2014pla}
and beyond~\cite{DelDuca:2019tur}.  Three-point form factors can also
exhibit factorized Regge behavior~\cite{Basso:2024hlx}. On the other hand,
the multi-Regge behavior of higher-point form factors is still ripe for
exploration.

In the planar limit of a large number of colors, $\mathcal{N}=4$ SYM
exhibits remarkable properties, such as the amplitude/Wilson loop
duality~\cite{%
Alday:2007hr,Drummond:2007aua,Brandhuber:2007yx,Drummond:2007au,
Alday:2008yw,Adamo:2011pv,Brandhuber:2010ad,Alday:2007he,Bern:2008ap,
Drummond:2008aq,Maldacena:2010kp,Ben-Israel:2018ckc,Bianchi:2018rrj}
and dual conformal invariance~\cite{%
Drummond:2006rz,Bern:2006ew,Bern:2007ct,Alday:2007hr,Drummond:2008vq}.
Such discoveries are often fueled by perturbative results at high
multiplicity and loop orders.

Recently, a novel antipodal duality has emerged from bootstrapped amplitudes
and form factors. The six-gluon maximally-helicity-violating (MHV) scattering
amplitude had been bootstrapped to seven loops~\cite{Caron-Huot:2019vjl}.
The three-point form factor for the chiral stress-tensor operator
(or equivalently $\tr\phi^2$) was then bootstrapped to eight
loops~\cite{Dixon:2020bbt,Dixon:2022rse}.
These two quantities turn out to be related by
{\it antipodal duality}~\cite{Dixon:2021tdw}: the combined action
of a kinematic map and reversing the order of letters in all terms
in the symbol.
The kinematic map maps the two-parameter phase space for the
form factor into a parity-preserving slice of the three-parameter phase space
for the six-point amplitude.
This duality can be checked to hold beyond the symbol, at least
modulo $i\pi$ terms which are not specified by the antipodal action
on the Hopf algebra for multiple polylogarithms~\cite{Dixon:2021tdw}.

More recently, the \emph{four}-point form factor for the same
$\tr\phi^2$ operator has been shown to exhibit antipodal
{\it self}-duality on a four-dimensional
(parity-preserving) slice of its five-dimensional phase
space~\cite{Dixon:2022xqh}.  This self-duality encompasses
the previous duality,
in the sense that the kinematic map in the four-point form factor
relates two different limits, one which produces the three-point form factor,
and the other produces the six-gluon scattering amplitude
(because it is also a triple-collinear splitting amplitude).
On the other hand, there is no fundamental understanding of
why antipodal {\it self}-duality should exist, nor has it been
verified yet beyond the symbol level.

Indeed, many high-order and high-multiplicity results to date are limited
to the symbol level, which leaves much beyond-the-symbol information
and numerical behavior unexplored.
Efforts have been made to recover the full function level information
for six-point amplitudes~\cite{%
Dixon:2013eka,Dixon:2014voa,Dixon:2014iba,Dixon:2015iva,
Caron-Huot:2016owq,Caron-Huot:2019vjl}
and, more recently, seven-point scattering amplitudes~\cite{Dixon:2020cnr}.
In fact, antipodal duality has been exploited to obtain the MHV six-point
amplitude at eight loops at function level~\cite{Dixon:2023kop}.

The three-point $\tr\phi^2$ form factor~\cite{Dixon:2020bbt,Dixon:2022rse}
has also been fixed at function level through eight loops.
However, beyond one loop, the knowledge of the four-point $\tr\phi^2$
form factor is presently limited to symbol level~\cite{Dixon:2022xqh} and
a special limit in which the operator has a light-like
momentum~\cite{Guo:2022qgv}.
The four-point $\tr\phi^3$ form factor, on the other hand,
involves simpler Feynman integrals than $\tr\phi^2$;
it has been computed at two loops using a bootstrap based on
master integrals~\cite{Guo:2021bym}.
(The three-point $\tr\phi^3$ form factor was very recently computed
through six loops~\cite{Basso:2024hlx,Henn:2024pki}.)

One difficulty in lifting the four-point $\tr\phi^2$ form factor to a
function is the large and intricate symbol alphabet:
The alphabet for the three-point $\tr\phi^2$
form factor has only 6 letters, while that for the six-point and seven-point
amplitude have 9 and 42 letters,
respectively~\cite{Goncharov:2010jf,Golden:2013xva}.
The symbol alphabet for
the $\tr\phi^2$ form factor at two (three) loops has
34 (88) letters~\cite{Dixon:2022xqh}.
The latter is considerably larger, and harder to rationalize,
than the seven-point amplitude's alphabet.

Lifting the symbol of the four-point form factor to a full function
would bring several new pieces of information.  The numerical values,
which are unavailable from only the symbol, could (if evaluated to high
enough loop order) probe the radius of convergence of perturbation
theory~\cite{Dixon:2014voa,Dixon:2015iva,Caron-Huot:2019vjl}.
It might also be possible to interpolate or extrapolate to 
strong coupling where a minimal surface formulation is
available~\cite{Alday:2007hr,Alday:2009dv,Maldacena:2010kp}.
In the case of amplitudes, such interpolation is relatively simple
at special kinematic points called origins~\cite{Basso:2020xts,Basso:2022ruw},
but it is also possible in the OPE limit~\cite{Basso:2013vsa,Sever:2020jjx}.
Antipodal self-duality can be analyzed beyond symbol level.
Various Minkowski factorization limits can also be studied,
such as multi-Regge kinematics,
the self-crossing (or pseudo-double-parton scattering)
limit~\cite{Georgiou:2009mp,Dixon:2016epj,Caron-Huot:2019vjl},
and the light-like limit~\cite{Guo:2022qgv}.

In this paper we uplift the two-loop four-point MHV form factor for
$\tr\phi^2$ from its symbol to a full function of the kinematics.
We do so by specifying
the iterated coproducts of the function (essentially its derivatives)
in a space of polylogarithmic functions with weight up to three,
as well as specifying their boundary values at a particular point in
the phase space.
We make special use of a three-parameter subspace of the five-parameter
phase space where all the symbol letters rationalize, which we call
the {\it rational surface}.  We write the form factor remainder explicitly
in terms of multiple polylogarithms on this surface, and use
that representation to move from region to region on the surface.

We fix beyond-the-symbol constants (which are zeta values)
using invariance under the dihedral symmetry group $D_4$,
and using the universal factorization behavior
in kinematic limits, including where a particle becomes soft,
or two or three particles become collinear.
We also match the near-collinear limit to data
from the FFOPE~\cite{Sever:2020jjx,Sever:2021nsq,Sever:2021xga,Basso:2023bwv}.

Very recently, the four-point $\tr\phi^2$ form factor has also been computed
at two loops~\cite{Guo:2024bsd} at function level,
by using unitarity cut methods to obtain
the loop integrands in $D$ spacetime dimensions.  Integration-by-parts
reduction was then used to write the result in terms of the basis
of two-loop, non-planar five-point master integrals with one external mass
provided in ref.~\cite{Abreu:2023rco}.  These master integrals are provided
in a $2\to3$ scattering configuration.  Ref.~\cite{Guo:2024bsd} also used
{\sc AMFlow}~\cite{Liu:2022chg} to compute a couple of numerical values
in (pseudo-)Euclidean or $1\to4$ decay kinematics.
As we focus mainly on the Euclidean region in our paper,
our results are complementary to those of ref.~\cite{Guo:2024bsd}.

This paper is organized as follows.  In Sec.~\ref{sec:R42} we introduce
our notation and review basic properties of form factors and
multiple polylogarithms. In Sec.~\ref{sec:uplift} we define the space of
functions used to describe the form factor, and the three-parameter rational
surface kinematics.  We then fix all the beyond-the-symbol constants
by utilizing the available information.
We provide the result for the remainder function in the bulk in
Sec.~\ref{sec:bulk}.  We conclude in Sec.~\ref{sec:concl} with a discussion
of future research directions enabled by this work.

Many of the explicit results of this paper are rather lengthy, so
they are included as computer-readable ancillary files:
{\tt AntipodalAlphabet.m} specifies the symbol alphabet we work with.
{\tt R42funcCoTable.m} gives the iterated coproducts of the function
in terms of a set of independent lower-weight functions we call
$P$ functions.
{\tt R42\_rational.m} is the representation of the remainder function on
the rational surface in terms of multiple polylogarithms.
{\tt Prat.m} provides the same representation for the $P$ functions.
{\tt PTT2to0\_xy.m} provides the $P$ functions in the OPE limit
(defined in Sec.~\ref{sec:uplift}).
{\tt R42\_OPE.txt} provides the set of coefficient functions describing
the remainder function in the OPE limit (see Appendix~\ref{app:nff}). \verb|R42_2d.m| provides the remainder function on the 2D kinematics (see Appendix \ref{app:2dkin}).


\section{Four-particle form factor and generalized polylogarithms}
\label{sec:R42}

\subsection{BDS-like normalized form factors}

In this paper, we study the MHV form factor for the chiral stress energy
tensor in planar $\mathcal{N}=4$ SYM.  One representative for this
BPS-protected operator super-multiplet is $\tr\phi^2$, where
$\phi^2$ is some traceless (non-Konishi) scalar bilinear, and the trace
``$\tr$'' is over the large-$N_c$ $SU(N_c)$ gauge group.
One component of the super four-point form factor is the matrix element
of $\tr\phi^2$ with two massless scalars and two same-helicity gluons:
\be
 \mathcal{F}^{\rm MHV}_4
 = \langle \tr\phi^2(q)\, \phi(p_1) \phi(p_2) g^+(p_3) g^+(p_4) \rangle \,.
\label{F4definition}
\ee
This form factor was first computed at one loop~\cite{Brandhuber:2010ad},
and then at two loops at symbol level~\cite{Dixon:2022xqh}.  Our task is
to provide a function-level description.

The operator momentum $q^\mu$ is the sum of the momenta of the four massless
particles,
\be
q^\mu = \sum_{i=1}^4 p_i^\mu \,,
\label{qdef}
\ee
where $p_i^2 = 0$.
The form factor depends on the external momenta $p_i$ through the
dimensionless ratios,
\be
  u_{i} \equiv \frac{(p_i+p_{i+1})^2}{q^2} \,,
  \qquad v_{i} \equiv \frac{(p_i+p_{i+1}+p_{i+2})^2}{q^2} \,,
\ee
where $i=1,2,3,4$ and all indices are mod 4.
These eight dimensionless ratios are constrained by three relations
from the masslessness of the $p_i$ and momentum conservation:
\begin{align}
-u_1 + u_3 + v_4 + v_1 &= 1  \, , \label{eq:uv_constraint_1} \\
-u_2 + u_4 + v_1 + v_2 &= 1 \, , \label{eq:uv_constraint_2} \\
-u_3 + u_1 + v_2 + v_3 &= 1 \, . \label{eq:uv_constraint_3}
\end{align}
There is a $D_4$ dihedral symmetry, which is generated by
two transformations,
\bea
&&{\bf cycle}\ ({\cal C}):\ \ p_i\to p_{i+1} \quad\Rightarrow\quad
u_i \to u_{i+1} \,, \ \ v_i \to v_{i+1} \,,
\label{eq:cyc}\\
&&{\bf flip}\ ({\cal F}):\ \ \ \ p_2 \lr p_4 \quad\Rightarrow\quad
u_1 \lr u_4 \,,\ \ u_2 \lr u_3 \,, \ \ v_1 \lr v_3 \,.
\label{eq:flip}
\eea

Following ref.~\cite{Dixon:2022xqh}, we normalize the MHV form
factor~(\ref{F4definition}) by a
BDS-like form factor which only depends on two-particle Lorentz invariants.
We define the function $\EE_4$ by
\be
 \mathcal{F}^{\rm MHV}_4 = 
 \mathcal{F}^{\rm MHV,tree}_4 \times
 \exp{\left[- \frac{\Gcusp(g^2)}{4\,\epsilon^2}
      \sum_{i=1}^4 \left( \frac{\mu^2}{-s_{i,i+1}} \right)^\epsilon \right]}
 \times \EE_4  \,,
\label{E4def}
\ee
where the 't Hooft coupling is $g^2 = N_c g_{\rm YM}^2/(16\pi^2)$ and
the cusp anomalous dimension in planar ${\cal N}=4$ SYM
is~\cite{Beisert:2006ez}
\be
\Gcusp(g^2)  = 4 g^2 - 8 \zeta_2 g^4 + 88 \zeta_4 g^6
- 4\left[ 219 \zeta_6 + 8 (\zeta_3)^2 \right] g^8 + \dots\,.
\ee

We define the {\it remainder function} $\RR_4$ by dividing
by the exponential of the full one-loop form factor (and taking the logarithm).
It is related to $\EE_4$ by
\be
\EE_4 = \exp{\left[ \frac{\Gamma_{\rm cusp}}{4} {\EE_4^{(1)}}
  + \RR_4 \right]} \,,
\ee
where the one-loop function $\EE_4^{(1)}$ is the finite part of the
one-loop amplitude, and is given by
\be
\EE_4^{(1)} =
- 2 \, \Li_2\bigl(1-v_1\bigr) - \Li_2\Bigl(1-\frac{u_2}{v_1v_2}\Bigr)
- \ln u_1 \ln u_2 + \ln v_1 \ln\Bigl(\frac{u_1u_2}{v_1v_2}\Bigr)
+ \zeta_2 \ +\ {\rm cyclic}.
\label{E41}
\ee
Here ``$+$~cyclic'' means to add the three images under the cyclic
transformation ${\cal C}$ in \eqn{eq:cyc}.

The form factor also has a representation in terms of a light-like polygonal
Wilson loop, which only closes in a space that is periodic by $q$,
in order to account for the operator momentum~\cite{Brandhuber:2010ad}.
In order to define a finite periodic polygonal Wilson loop,
one can normalize by suitable lower-point quantities, resulting
in a ``framed'' Wilson loop
$\mathcal{W}_4$~\cite{Sever:2020jjx,Sever:2021nsq,Sever:2021xga}.
The framed Wilson loop is related to the form-factor remainder function by
\be
\mathcal{W}_4 = \exp{\left[
\frac{\Gamma_{\rm cusp}}{4} \mathcal{W}_4^{(1)} + \RR_4 \right]} \,,
\label{Wdef}
\ee
where
\bea
\mathcal{W}_4^{(1)} &=& \EE_4^{(1)}
+ \ln^2(v_1v_4-u_1) + 2 \ln^2(1-v_4) + 2 \ln^2 v_4
\nn\\&&\hskip0cm\null
- \ln(v_1v_4-u_1) [ 2 \ln v_4 - 2 \ln(1-v_4)
                 + \ln u_1 + \ln u_2 + \ln u_3 - \ln u_4 ]
\nn\\&&\hskip0cm\null
- 2 \ln(1-v_4) [ 2 \ln v_4 + \ln u_3 ]
+ 2 \ln v_4  [ \ln u_2 + \ln u_3 - \ln u_4 ]
\nn\\&&\hskip0cm\null
+ \ln u_1 \ln u_2 - \ln u_2 \ln u_4 + \ln u_4 \ln u_1 + 2 \zeta_2 \,.
\label{W41}
\eea

Due to the correspondence between the form factor and periodic Wilson loops,
we can parametrize the kinematics by the
coordinates $\tau_i, \sigma_i, \phi_i$ used in the
FFOPE~\cite{Sever:2020jjx,Sever:2021nsq,Sever:2021xga}.
We define
\be
    T = e^{-\tau}, \quad S = e^{\sigma},\quad T_2 = e^{-\tau_2}, \quad
    S_2 = e^{\sigma_2},\quad F_2 = e^{i\phi_2}.
\ee
The dimensionless ratios $u_i, v_i$ are related to $T, S, T_2, S_2, F_2$ by
\begin{align}
  u_1&= \frac{T^2 T_2^2}{\left(T^2+1\right) \left(S^2+T^2+T_2^2+1\right)} \,,
\nonumber\\
  u_2&= \biggl[ 1 + T^2
    + \frac{ S^2 [ S_2 T_2 (1 + F_2^2 )
                + F_2 (1 + S_2^2 + T^2 + T_2^2) ] }{F_2 S_2^2}\biggr]^{-1}  \,,
\nonumber \\
  u_3&= \frac{S^2}{\left(T^2+1\right) \left(S^2+T^2+T_2^2+1\right)} \,,
\label{eq:ope_parametrization} \\  
  u_4&= \frac{S^2 T^2}{S_2^2} u_2 \,,  \nonumber \\
  v_1&= \frac{T_2^2+1}{S^2+T^2+T_2^2+1} \,, \nonumber\\
  v_2&= 1 + u_2 - u_4 - v_1 \,, \nonumber\\
  v_3&= 1 - u_1 + u_3 - v_2 \,, \nonumber\\
  v_4&= \frac{T^2}{T^2+1} \,. \nonumber
\end{align}

\subsection{Polylogarithms and (antipodable-)symbol alphabet}

We expand the remainder function perturbatively as
\be
    \RR_4 = \sum_{L=2}^\infty g^{2L} \RR^{(L)}_4 \,,
\ee
and similarly for the function $\EE_4$.
The $L$-loop quantities $\RR^{(L)}_4$ and $\EE^{(L)}_4$
are expected to be multiple polylogarithms of weight $2L$.

Multiple polylogarithms (MPLs) are iterated integrals over a logarithmic
kernel~\cite{Chen, Gonch3, Goncharov:2010jf, Duhr:2011zq, Duhr:2012fh}.
The total differential of a weight $n$ MPL has the form
\be \label{eq:total_diff}
dF = \sum_{\phi\in\Phi} F^\phi \, d \ln \phi \, ,
\ee
where the sum is over {\it letters} $\phi$ which belong to the
{\it symbol alphabet} $\Phi$, and the $\phi$-{\it coproducts} $F^\phi$
appearing in \eqn{eq:total_diff} have weight $n-1$.
In integral form, MPLs are commonly expressed as $G$ functions,
\be \label{eq:G_func_def}
G_{a_1,a_2,\ldots,a_n}(z) = \int_0^z \frac{dt}{t-a_1} G_{a_2,\dots, a_n}(t)\,,
\qquad G_{\fwboxL{27pt}{{\underbrace{0,\dots,0}_{p}}}}(z) = \frac{\ln^p z}{p!} \,.
\ee

The classical polylogarithms $\Li_n$ are special cases of $G$ functions,
\be
\Li_n(z) = -G_{\fwboxL{34pt}{{\underbrace{0,\dots,0}_{n-1},1}}}(z) \, .
\ee
Harmonic polylogarithms (HPLs)~\cite{Remiddi:1999ew} $H_{\vec{a}}(z)$ with $a_i \in \{0,1,-1\}$ are another special case, with
\be
 H_{\vec{a}}(z)=(-1)^p G_{\vec{a}}(z)\,,
\ee
where $p$ is the number of `$1$'s in the index list $\vec{a}$.
Transcendental constants such as multiple zeta values (MZVs)
can be viewed as special values of the functions~\eqref{eq:G_func_def}.

The weight is the number of logarithmic integrations that appear;
the weight of a product of polylogarithms is given by the sum of weights
of the factors in the product.  In the $G_{a_1,\dots, a_n}(z)$ notation,
the weight simply corresponds to the number of indices $n$.

The symbol~\cite{Goncharov:2010jf}
of a generic polylogarithmic function $F$ is defined recursively
in terms of its total differential~\eqref{eq:total_diff}:
\be
\mathcal{S} \left(F  \right) = \sum_\phi \mathcal{S}(F^\phi) \otimes \phi \, ,
\label{SymbolDef}
\ee
where $\mathcal{S}(\ln\phi) = \phi$ for letters $\phi$ by convention.
Formally the symbol is also the maximal iteration of the (motivic)
coaction $\Delta$ associated with a Hopf algebra for
MPLs {(up to constant entries of the
coproduct such as $\ln 2$)}~\cite{%
Gonch3,FBThesis,Goncharov:2010jf,Brown1102.1312,
Duhr:2012fh,2015arXiv151206410B}.

The derivatives~(\ref{eq:total_diff}) are smooth wherever all
the letters in the symbol alphabet $\Phi$ are nonvanishing.
Conversely, the vanishing loci of the letters $\phi$ that appear in the
tensor product~(\ref{SymbolDef}) provide the locations of possible branch cuts.
While the first derivatives are encoded in the last
entry of the symbol, according to the iterative definition~(\ref{SymbolDef}),
branch cuts can be taken by clipping off suitable first entries.
Not all branch cuts are allowed singularities on physical sheets. Requiring
only branch cuts at physical locations
imposes restrictions on the symbol, called first-entry conditions,
and it implies additional restrictions on the function.  

The symbol of the four-point $\tr\phi^2$ form factor was first bootstrapped
at two loops~\cite{Dixon:2022xqh}
by starting with a list of 113 symbol letters collected from all the relevant
two-loop one-mass five-point
integrals~\cite{Abreu:2020jxa,Abreu:2021smk,Abreu:2023rco}.
{(Note that ``planar'' ${\cal N}=4$ SYM refers to the leading-color approximation. Non-planar Feynman diagrams can contribute to leading-color form factors of color-singlet operators -- provided that the diagram becomes planar when one deletes the external leg corresponding to the operator.)}
The two-loop symbol only requires 34 out of the 113 letters.
The three-loop symbol has been bootstrapped successfully~{\cite{Dixon:2022xqh}} with the assumption that no new letters arise at three loops.  It requires 88 out of the 113 letters. If we assume that no new letters arise at higher loops, and we also assume antipodal self-duality at the symbol level, then we find an allowed symbol alphabet of 93 letters.  The other 20 of the 113 letters map, under the kinematic map, to functions that are outside of the 113-letter alphabet~\cite{Gurdoganprivate}.

This 93-letter ``antipodal'' alphabet is described in the
ancillary file {\tt AntipodalAlphabet.m}.
The alphabet features five square roots $\sqrt{\Delta_{1a}}$, $\sqrt{\Delta_{1b}}$,
$\sqrt{\Delta_{2a}}$, $\sqrt{\Delta_{2b}}$, and $\sqrt{\Delta_{3}}$, where
\be
\begin{aligned}
    {\Delta_{1a}} & = (v_1 + v_2)^2 - 4 {u_2}, \\ 
 {\Delta_{1b}} & = 
 1 - 2 {u_1} + {u_1}^2 - 2 {u_3} - 2 {u_1} {u_3} + {u_3}^2, \\ 
 {\Delta_{2a}} & = 
 {u_1}^2 - 2 {u_1} {v_1} + {v_1}^2 + 2 {u_1}^2 {u_2} - 2 {u_1} {v_1} {u_2} + 
  {u_1}^2 {u_2}^2 - 2 {u_1}^2 {v_2} + 4 {u_1} {v_1} {v_2} - 2 {v_1}^2 {v_2} \\ & - 
  2 {u_1}^2 {u_2} {v_2} + 2 {u_1} {v_1} {u_2} {v_2} + {u_1}^2 {v_2}^2 - 
  2 {u_1} {v_1} {v_2}^2 + {v_1}^2 {v_2}^2 - 2 {u_1} {v_1} {u_3} + 2 {v_1}^2 {u_3} - 
  2 {u_1} {u_2} {u_3} \\ &- 2 {v_1} {u_2} {u_3} + 2 {u_1} {v_1} {u_2} {u_3} - 
  2 {u_1} {u_2}^2 {u_3} + 2 {u_1} {v_1} {v_2} {u_3} - 2 {v_1}^2 {v_2} {u_3} + 
  2 {u_1} {u_2} {v_2} {u_3} + 2 {v_1} {u_2} {v_2} {u_3} \\ &  + {v_1}^2 {u_3}^2 - 
  2 {v_1} {u_2} {u_3}^2 + {u_2}^2 {u_3}^2, \\ 
 {\Delta_{2b}} & = 
 {u_1}^2 {u_2}^2 - 2 {u_1} {u_2} {v_2} - 2 {u_1}^2 {u_2} {v_2} + 
  2 {u_1} {v_1} {u_2} {v_2} + {v_2}^2 + 2 {u_1} {v_2}^2 + {u_1}^2 {v_2}^2 - 
  2 {v_1} {v_2}^2 \\ & - 2 {u_1} {v_1} {v_2}^2 + {v_1}^2 {v_2}^2 - 2 {u_1} {u_2} {u_3} + 
  2 {u_1} {v_1} {u_2} {u_3} - 2 {u_1} {u_2}^2 {u_3} - 2 {v_2} {u_3} - 
  2 {u_1} {v_2} {u_3} + 4 {v_1} {v_2} {u_3} \\ & + 2 {u_1} {v_1} {v_2} {u_3} - 
  2 {v_1}^2 {v_2} {u_3} - 2 {u_2} {v_2} {u_3} + 2 {u_1} {u_2} {v_2} {u_3} + 
  2 {v_1} {u_2} {v_2} {u_3} + {u_3}^2 - 2 {v_1} {u_3}^2 + {v_1}^2 {u_3}^2 \\ & + 
  2 {u_2} {u_3}^2 - 2 {v_1} {u_2} {u_3}^2  + {u_2}^2 {u_3}^2, \\ 
 {\Delta_{3}} & = 
 {u_2}^2 - 2 {u_1} {u_2}^2 + {u_1}^2 {u_2}^2 + 2 {u_1} {u_2} {v_2} - 
  2 {u_1}^2 {u_2} {v_2} - 2 {v_1} {u_2} {v_2} + 2 {u_1} {v_1} {u_2} {v_2} + 
  {u_1}^2 {v_2}^2 \\ & - 2 {u_1} {v_1} {v_2}^2 + {v_1}^2 {v_2}^2 - 4 {u_1} {u_2} {u_3} + 
  2 {v_1} {u_2} {u_3} + 2 {u_1} {v_1} {u_2} {u_3} - 2 {u_2}^2 {u_3} - 
  2 {u_1} {u_2}^2 {u_3} \\ & + 2 {u_1} {v_1} {v_2} {u_3} - 2 {v_1}^2 {v_2} {u_3} + 
  2 {u_1} {u_2} {v_2} {u_3} + 2 {v_1} {u_2} {v_2} {u_3} + {v_1}^2 {u_3}^2 - 
  2 {v_1} {u_2} {u_3}^2 + {u_2}^2 {u_3}^2 \,.
\end{aligned}
\ee
Of the 93 letters, 56 are rational.  The other 37 have the form
$(a_j+\sqrt{\Delta_m})/(a_j-\sqrt{\Delta_m})$ for some polynomials
$a_j(u_i,v_i)$ and $m=1a,2a,1b,2b,3$; these 37 letters are odd under some of
the Galois symmetries that flip the signs of the various square roots.
The form factor should be even under all such Galois symmetries.
Flipping the sign of $\sqrt{\Delta_3}$ corresponds to spacetime parity.
Antipodal self-duality holds on the parity-preserving surface $\Delta_3=0$,
which is $F_2=1$ in the parametrization~(\ref{eq:ope_parametrization}).

Although a complete functional description was already given
for the relevant two-loop integrals~\cite{Abreu:2023rco}, it was
given (so far) only in the kinematical region for $2 \to 3$ scattering
where the massive leg has positive mass and is on the outgoing side.
We will be interested in other kinematical regions. Also, we wish to
extend the description (eventually) to higher loop functions
with the same symbol alphabet. Therefore, in the following we will provide
an alternate functional description, valid at least for various subspaces
of the kinematics.  We will focus more on the Euclidean region, which
also is relevant for the pseudo-Euclidean region of $1\to4$ decay kinematics,
where the operator is massive and in the initial state. It would be interesting
to connect to the description in
refs.~\cite{Abreu:2023rco,Guo:2024bsd} in future work.


\section{Integrating the symbol up to functions }
\label{sec:uplift}

The two-loop four-gluon form factor has been bootstrapped at the symbol
level~\cite{Dixon:2022xqh}.  At this level, all MZVs vanish,
$\mathcal{S}({\rm MZV}) = 0$.  In order to describe the form factor at the
function level, we need to recover the MZVs.  We can do this iteratively
in the differential definition~(\ref{eq:total_diff}),
by providing function-level
values for the single coproducts $F^\phi$ and/or multiple coproducts
-- which we generically call $P$ \emph{functions} --
and by providing boundary values for these quantities at specific points.
On specific surfaces, we can integrate all the way up to the weight 4
$\RR_4^{(2)}$ in terms of $G$ functions (or HPLs, for simple enough
surfaces).


\subsection{Function space}

The framework for integrating up the remainder function from lower-weight
functions is the coproduct formalism~\cite{Dixon:2013eka,Caron-Huot:2019bsq}.
We construct a set of basis functions, $\{F^{(w)}_i\}$, for each
weight $w$ up to $2L$.  The sets have dimensions $|F^{(w)}|$ and are big enough
to contain all the (multiple) coproducts of the form factor or amplitudes.
The different bases are linked to each other by the coproducts $\Delta_{w-1,1}$,
which connect two consecutive bases, for weights $w-1$ and $w$,
via a three-index tensor $T^{(w)}_{ij\phi}$:
\be
\Delta_{w-1,1} F^{(w)}_i = \sum_{j,\phi} T^{(w)}_{ij\phi} F^{(w-1)}_j \otimes \phi \,.
\label{Tdef}
\ee
Here $i,j$ are indices labelling the basis functions at weight $w$ and $w-1$,
respectively;
$\phi$ are letters in the symbol alphabet $\Phi$;
$T^{(w)}_{ij\phi}$ are rational numbers filling out a three-index tensor with
dimension $|F^{(w)}|\times|F^{(w-1)}|\times |\Phi|$.
The symbol-level information corresponds to the coefficients of functions
with nonvanishing symbols, whereas the coefficients of MZVs within the
$\{F^{(w)}_i\}$ are yet to be fixed.  Generally the dimensions $|F^{(w)}|$
have to increase from the symbol-level version, in order to accommodate the MZVs.
In our two-loop case, the increase will be very modest, just an increase of
one function at weight 2, to account for $\zeta_2$.

This coproduct table effectively defines an iterative differential equation
for each basis element,
\be
\label{eq:differential}
    dF_i^{(w)} = \sum_{j,\phi} T^{(w)}_{ij\phi} F_j^{(w-1)} d\ln \phi.
\ee
Once we know the value at any particular point, with the function-level
coproduct table, we will be able to integrate up the functions (at least
numerically) at an arbitrary kinematic point.

For the description of $\RR_4^{(2)}$ we choose a minimal space of
$P$ functions with weight up to 4, based on the symbol-level information.
At weight 1 we can only have 8 independent functions $\{ \ln u_i$, $\ln v_i \}$,
because only these logarithms have branch cuts in the correct location,
i.e.~the first-entry condition.
{(We remark that eqs.~\eqref{eq:uv_constraint_1}--\eqref{eq:uv_constraint_3} are constraints on the underlying kinematic variables, which do not imply any linear dependence of their logarithms, the symbol letters.)}
As mentioned above, only 34 of the 93 antipodal letters appear in the
two-loop symbol~\cite{Dixon:2022xqh}.  By taking the iterated $\Delta_{n-1,1}$
coproducts of the symbol $\RR_4^{(2)}$, we find that there are 9
independent single coproducts at weight 3 ($\{3,1\}$ coproducts)
and 32 independent double coproducts ($\{2,1,1\}$ coproducts).
Therefore at symbol level the set of symbols that we need to upgrade
to functions consists of 32 at weight two,
9 at weight three, plus 1 weight-four function, $\RR_4^{(2)}$ itself.

To fix the beyond-the-symbol constants we make use of the following constraints:
\begin{enumerate}
\item {\bf Integrability}: An integrable function must have commuting partial
    derivatives. Thus if we apply the differential \eqref{eq:differential} twice
    we need (for variables $x,y \in {u_i,v_i}$):
\be
\frac{\partial^2 F}{\partial x \partial y}
 = \frac{\partial^2 F}{\partial y \partial x} \,.
\ee
This condition results in a large set of 3774 independent linear
relations among the $93\times 93$ double coproducts $F^{\phi_i,\phi_j}$,
which have to hold at function level too.
\item {\bf Physical branch cuts}: The form factor can only develop logarithmic
divergences at physical branch points.  On the Euclidean sheet, these
singularities occur only where the Mandelstam variables
$s_{ij}$ or $s_{ijk}$ vanish, or equivalently where $u_i$ or $v_i\to0$.
In other words, any function $F$ in the space has to be nonsingular as
$\phi \to 0$ for any letter that is not in the first-entry,
$\phi \notin \{u_i,v_i\}$. This condition constrains the first coproducts in
particular limits, because derivatives in the singular direction must vanish
as $\phi \to 0$.
\item {\bf Extended Steinmann relations}: The BDS-like normalized form factor
$\EE_4$ should respect the Steinmann relations. The double discontinuity
associated with cuts in two overlapping 3-particle channels labeled by
$s_{i,i+1,i+2}$ must vanish~\cite{Dixon:2022xqh}.
(Note that the exponential factor that is removed in \eqn{E4def} contains
only two-particle invariants.)
In terms of the dimensionless letters $v_i = s_{i,i+1,i+2}/q^2$, the
following double coproducts vanish
\be
F^{v_i,v_j} = 0, \quad i\neq j \,.
\label{ESrel}
\ee
for the BDS-like normalized function $\EE_4$, or any of its coproducts.
Strictly speaking, the Steinmann relations only imply \eqn{ESrel} in the
first two slots, i.e.~for the weight 2 space $F^{(2)}$.  However, in practice
we find that this adjacency relation can be applied to all symbol entries in
the middle in the two-loop case (but not for $v_i,v_{i+2}$ in the
three-loop case).
So we could in principle apply $F^{v_i,v_{i+1}} = 0$, everywhere also
in the lower-weight functions.  However, on the rational surface we use
for fixing constants (see Sec.~\ref{subsubsec:rational}), the limit
$u_2 \to v_1 v_2$ means that the association
of $v_1$ and $v_2$ with 3-particle channels is obscured, and effectively 
only $F^{v_3,v_4} = 0$ can be used. 
\item {\bf Cycle and flip symmetry}: $\EE_4$ and $\RR_4$ are
 invariant under the dihedral group $D_4$, which is generated by the
 cycle and flip transformations in eqs.~(\ref{eq:cyc}) and (\ref{eq:flip}).
\item {\bf Factorization limits}:  As described in more detail below,
we use universal factorization behavior in soft and collinear kinematic limits,
and we match the near-collinear limit to data from the FFOPE.
\end{enumerate}

Taking into account all this information, we are able to fix all the MZVs.
We find that $\zeta_2$ needs to be added as an independent
weight-2 function, while $\zeta_3$ can be absorbed into the existing 9 weight
3 symbol-level coproducts.  Therefore the spaces of functions needed to
describe $\RR_4^{(2)}$ have dimensions $|F^{(1)}| = 8$,  $|F^{(2)}| = 33$,
$|F^{(3)}| = 9$,  $|F^{(4)}| = 1$. Our result for the iterated coproduct table
at function level is contained in the ancillary file {\tt R42funcCoTable.m}.

\subsection{Boundary kinematics}

\begin{figure}
\centering
\includegraphics[height=8cm]{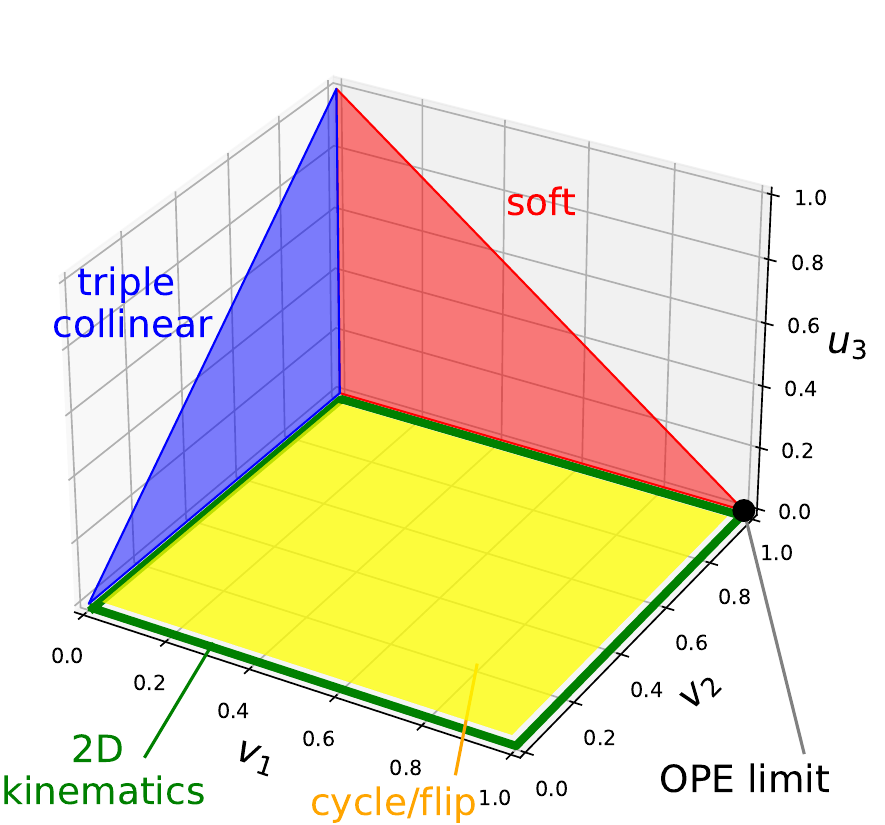}
    
\caption{ The rational surface is parametrized by three kinematic variables
$(u_3,v_1,v_2)$.  It intersects a soft limit at the $v_2 = 1$ surface (red).
The intersection with a triple collinear limit is at the $v_1=0$ surface (blue).
The $u_3=0$ surface (yellow) maps into itself under two dihedral transformations
${\cal C}^2$ and ${\cal C}\cdot{\cal F}$, as indicated by ``cycle/flip''.
The (green) square boundary of the $u_3=0$ surface intersects another
two-parameter surface~(\ref{2Dkin})
where the momenta all lie in two spacetime dimensions.
The point $(0,1,1)$ makes contact with the OPE limit. }
    \label{fig:rat3d}
\end{figure}

We first consider a three-parameter rational surface
where the symbol alphabet simplifies so that all letters are rational
functions of $u_3,v_1,v_2$.  This surface also interpolates between
soft, collinear, OPE, multi-Regge, and self-crossing limits.
Using these limits, we can deduce
the $G$-function representation of the remainder function
on the rational surface, $\RR_4^{(2)}(u_3,v_1,v_2)$.  
This representation automatically
gives zeta-valued information for coproducts corresponding
to derivatives in directions tangent to the rational surface.
However, other coproducts are needed for derivatives along directions
normal to the rational surface.
To determine them, we use bulk conditions such as integrability.
In this way, we can recover the function-level information for the
full coproduct table in the bulk.
Given the expressions for all the basis functions $F_i^{(w)}$
(the $P$ functions) on the rational surface boundary,
the function in the bulk is uniquely defined. 
We illustrate the rational surface, and how it connects
different kinematic limits, in Fig.~\ref{fig:rat3d}.

\subsubsection{Rational surface and soft/collinear factorization}
\label{subsubsec:rational}

The rational surface is defined by the limit
\be
\frac{u_2}{v_1 v_2} \to 1, \qquad u_1 \to 0,
\label{ratsurf}
\ee
with $u_3,v_1,v_2$ generic.  On this surface, the 93-letter antipodal
symbol alphabet simplifies to the following 20 letters:
\be
\begin{aligned}
\Phi_{\rm rational} =\ & \big\{    {u_3},\quad    {v_1},\quad    {v_2},\quad  
1 - {u_3},\quad    1 - {v_1},\quad    1 - {v_2},\quad    1 + {u_3},\quad  
{u_3} + {v_1}, \\
&    -{u_3} + {v_2},\quad    {v_1} - {v_2},\quad  1 - {u_3} - {v_1},
\quad    1 + {u_3} - {v_2},\quad    {u_3} + {v_1} - {v_2},\quad  \\
&-{u_3} + {v_2} - {v_1} {v_2},\quad    {u_3} + {v_1} - {v_1} {v_2},
 \quad   -{u_3} + {u_3} {v_1} + {v_2} - {v_1} {v_2} - {u_3} {v_1} {v_2},\quad  \\
&-{u_3} + {v_2} - 2 {v_1} {v_2} + {u_3} {v_1} {v_2} + {v_1^2} {v_2},\quad  
-{u_3} {v_1} - {u_3} {v_2} + {u_3} {v_1} {v_2} + {v_2^2} - {v_1} {v_2^2},\quad  \\
&\,\, {u_3} + {v_1} - 2 {v_1} {v_2} - {u_3} {v_1} {v_2} + {v_1} {v_2^2},\quad 
       {u_3} {v_1} + {v_1^2} + {u_3} {v_2} - {u_3} {v_1} {v_2} - {v_1^2} {v_2} \big\}.
\end{aligned}
\label{ratalphabet}
\ee
All the letters are now rational (polynomial) in $u_3,v_1,v_2$.
There is also the trivial infinitesimal letter $u_1$, which factors
out on the rational surface, i.e.~all functions are polynomial in $\ln u_1$.
Furthermore, at two loops the last four letters do not appear
and the alphabet becomes \textit{linearly reducible}~\cite{Duhr:2019tlz},
thus enabling us to conveniently represent $\RR_4^{(2)}$
in terms of $G$-functions in this limit.

We remark that at three loops, although three of the last four letters in
\eqn{ratalphabet} appear in the symbol, they never appear together in the same
symbol term.
This feature allows for some contributions to be linearized in $v_1$, and others
in $v_2$, so that a representation of the 3-loop remainder function in terms
of $G$-functions with rational arguments appears feasible as well.

{Given the symbol,} the function-level result on the rational surface can be fully determined
by the physical branch cut conditions, cycle-flip symmetry,
and soft/collinear limits, as we shall now discuss.

First, we discuss the branch-cut conditions.
In the rational alphabet~(\ref{ratalphabet})
the letters at two loops that do not correspond to physical Mandelstam
variables (taking into account $u_2=v_1v_2$, $u_4=(1-v_1)(1-v_2)$, etc.) are
\be
\begin{aligned}
& 1 - {u_3}, \quad   1 + {u_3}, \quad {u_3} + {v_1}, \quad  -{u_3} + {v_2},
\quad    {v_1} - {v_2}, \quad {u_3} + {v_1} - {v_2}, \quad  \\
&-{u_3} + {v_2} - {v_1} {v_2},\quad {u_3} + {v_1} - {v_1} {v_2},
\quad   -{u_3} + {u_3} {v_1} + {v_2} - {v_1} {v_2} - {u_3} {v_1} {v_2}.
\end{aligned}
\label{ratnotfirst}
\ee
These letters $\phi$ should not give rise to singularities of form factors.
That is, $\RR^{(2)}_4$ and its coproducts should be finite as $\phi\to0$,
where $\phi$ are the letters in \eqn{ratnotfirst}.
We have observed this at the symbol level, where the finiteness is
due to letters preceding $\phi$ going to 1 as $\phi\to0$, which
leads to power-law vanishing in this region.
At the function level, we need to remove all zeta-value containing
functions that have unphysical branch cuts, e.g. $\zeta_2 \ln\phi$
behavior for any coproduct $F$ as $\phi\to0$. So we require that
$F^\phi \to 0 \times \zeta_2$ in this region (and similarly with $\zeta_3$
at weight 3).

Secondly, we discuss consequences of dihedral symmetry.
The rational surface~(\ref{ratsurf})
can be related to another rational surface,
$\frac{u_2}{v_1 v_2} \to 1, u_3 \to 0$,
by a cycle-then-flip transformation,
\be
\label{eq_cyc_flip}
\mathcal{F}\cdot \mathcal{C}:\ \
p_1 \lr p_4,\ \ p_2 \lr p_3
\quad\Rightarrow\quad
u_1 \lr u_3,\ \ v_1 \lr v_2,\ \ v_3 \lr v_4 \,.
\ee
These two dihedral images of the rational surface make contact at a
sub-limit of the first surface, $u_3 \to 0$.
Therefore, the two-parameter surface parametrized by $v_1, v_2$ at
\be 
\frac{u_2}{v_1 v_2} \to 1,\ \ u_1 \to 0,\ \ u_3 \to 0
\label{ratu3to0}
\ee
is mapped to itself through the reflection~(\ref{eq_cyc_flip})
which exchanges $v_1 \lr v_2$.
This symmetry requires $\RR^{(2)}_4(u_3,v_1,v_2)$
to vanish at the rational surface boundary $u_3 \to 0$.

Thirdly, there are a few other sub-limits where the behavior of $\RR_4^{(2)}$
is known by soft-collinear factorization.
The soft limit of the external momentum $p_1 \to 0$ is
contained inside the rational surface as $v_2\to 1$.
In this soft limit, the four-point form-factor remainder function
goes smoothly into the three-point form factor,
\be
\RR_4^{(2)} |_{{\rm rational} , \, v_2 \to 1}\ \ \to\ \ \RR_3^{(2)} \,.
\label{soft1limit}
\ee
The right-hand side is known at function level
for general kinematics~\cite{Brandhuber:2012vm}:
\be
\begin{aligned}
\label{eq:3ptff}
\RR^{(2)}_3(u,v) =& -2 \left[
\mathrm{J}_4 \left( -\frac{u v}{w}\right)
+\mathrm{J}_4 \left( -\frac{v w}{u}\right)
+\mathrm{J}_4 \left( -\frac{w u}{v}\right)\right]
-8 \sum_{i=1}^3 \left[
\mathrm{Li}_4 \left(1-x_i^{-1}\right)+\frac{\log^4 x_i}{4!} \right] \\
& -2 \left[ \sum_{i=1}^3 \mathrm{Li}_2 (1-x_i^{-1}) \right]^2
+\frac{1}{2} \left[ \sum_{i=1}^3 \log^2 x_i\right]^2
- \frac{\log^4(u v w)}{4!} - \frac{23}{2} \zeta_4  \,,
\end{aligned}
\ee
with
\be
\mathrm{J}_4(z) = \mathrm{Li}_4(z)-\log(-z) \mathrm{Li}_3(z)
+\frac{\log^2(-z)}{2!} \mathrm{Li}_2(z)
-\frac{\log^3(-z)}{3!} \mathrm{Li}_1(z) - \frac{\log^4(-z)}{48} \,,
\ee
where $x_1=u = \frac{s_{12}}{s_{123}}$, $x_2=v= \frac{s_{23}}{s_{123}}$
and $x_3=w = 1-u-v = \frac{s_{31}}{s_{123}}$
are the dimensionless ratios parametrizing the three-point form factor.
They are related to the $u_i, v_i$ variables
in the four-point form factor in the soft limit by
\be
u = v_1,\quad v = u_3,\quad w = 1-v_1- u_3.
\ee
In other words, the rational surface soft limit is fixed at function level
by $\RR_4^{(2)}(u_3,v_1,1) = \RR_3^{(2)}(v_1,u_3) = \RR_3^{(2)}(u_3,v_1)$,
using also the $D_3$ dihedral symmetry of $\RR_3$.

Fourthly, another sub-kinematics inside the rational surface is the
triple collinear limit where three external momenta are parallel.
In OPE variables this corresponds to letting $T \to 0$,
for a parametrization like \eqn{eq:ope_parametrization} but after cycling
$u_i \to u_{i+1}$, $v_i \to v_{i+1}$.
The remainder function $\RR_4^{(2)}$ in this limit, at leading power in $T$,
reduces smoothly to the remainder function of the MHV six-gluon
scattering amplitude $R_6^{(2)}$~\cite{Dixon:2022xqh}.
This result follows from dual conformal invariance and
factorization~\cite{Bern:2008ap}, and it can also be seen
in the FFOPE framework~\cite{Sever:2020jjx,
Sever:2021nsq,Sever:2021xga,Basso:2023bwv}.
The three-parameter rational surface makes contact with the
triple collinear limit on a two-parameter surface at $v_1 \to 0$.
(Note that two of the two-particle invariants vanish as well in this
rational-surface limit, $u_1$ and $u_2 = v_1 v_2$.)
The six-point remainder function in this limit is:
\bea
\lim_{\hat{v} \to 1,\hat{w}\to 0} R^{(2)}_6(\hat{u},\hat{v},\hat{w})
&=& \ln \hat{w} \bigl[ 2 {\rm Li}_3(\hat{u})
                     - \ln \hat{u} {\rm Li}_2(\hat{u}) \bigr]
- 4 {\rm Li}_4(\hat{u}) + 2 {\rm Li}_4\left(\frac{-\hat{u}}{1-\hat{u}}\right)
- \frac{1}{2} \bigl[ {\rm Li}_2(\hat{u}) \bigr]^2
\nn\\ &&\hskip0cm\null
+ \ln \hat{u} \biggl[ 2 {\rm Li}_3(\hat{u})
            - \ln(1-\hat{u}) {\rm Li}_2(\hat{u})
            - \frac{1}{3} \ln^3(1-\hat{u}) \biggr]
+ \frac{1}{12} \ln^4(1-\hat{u})
\nn\\ &&\hskip0cm\null
+ \zeta_2 \bigl[ 2 {\rm Li}_2(\hat{u}) + \ln^2(1-\hat{u}) \bigr]  \,,
\label{eq:6pt_tricol}
\eea
where $\hat u = \frac{s_{12} s_{45}}{s_{123} s_{345}}$,
$\hat v = \frac{s_{23} s_{56}}{s_{234} s_{456}}$,
$\hat w = \frac{s_{34} s_{61}}{s_{345} s_{561}}$ are cross ratios
describing the six-gluon kinematics.
They are related to the $u_i, v_i$ describing the four-point form factor by 
\be 
\hat u = \frac{u_3 (1-v_2)}{v_2(1-u_3)},\quad
\hat w = \frac{u_1}{v_1 (1-u_3)}, \quad \text{with } u_1 \ll v_1 .
\label{uw_relation}
\ee
In summary, the triple-collinear boundary condition is
$\RR_4^{(2)}(u_3,v_1,v_2)|_{v_1\to0} = R_6^{(2)}(\hat{u},1,\hat{w})$,
where $\hat{u}, \hat{w}$ are given in \eqn{uw_relation}.

These constraints fully determine the remainder function on the
rational surface at function level.  On this surface, the remainder function
has a mild logarithmic singularity due to the fact that $u_1 \to 0$:
\be
\RR_4^{(2)}(u_1;u_3,v_1,v_2) = D_0(u_3,v_1,v_2) + D_1(u_3,v_1,v_2) \ln u_1 \,.
\label{D0D1decomposition}
\ee
The coefficient of $\ln u_1$ is a weight 3 MPL,
\be
\begin{aligned}
D_1 &= 4 \zeta_2 \Bigl[ G_{0}(1-v_1) + G_{0}(1-v_2) - G_{1}(v_2)
                 + G_{1-v_1}(u_3) - G_{0}(1-v_1-u_3) \Bigr] \\
& + \Bigr( G_{0}(v_2) - G_{1}(v_2) \Bigr) \Bigl[
    G_{-v_1,-1}(u_3) + G_{-v_1,1-v_1}(u_3) - G_{1-v_1,1-v_1}(u_3)
  + G_{v_2,v_2}(u_3) \\
& \hskip1cm
 + G_{v_2,-1+v_2}(u_3) - 2 G_{0,v_2}(u_3)
 - G_{0,-1}(u_3) + G_{1-v_1,v_2}(u_3) + G_{v_2,1-v_1}(u_3) \\
& \hskip1cm
 - G_{-v_1+v_2,1-v_1}(u_3)
 - G_{-v_1+v_2,-1+v_2}(u_3)  \\
& \hskip1cm
 + G_{0,(1-v_1)v_2/(1-v_1+v_1 v_2)}(u_3)
 - G_{v_2,(1-v_1)v_2/(1-v_1+v_1 v_2)}(u_3) \Bigr] \\
&
- 4 G_{0,1,0}(u_3)
+ 2 \Bigl[ G_{0,1-v_1,0}(u_3) + G_{1-v_1,1,0}(u_3)
      + G_{0,v_2,0}(u_3) + G_{v_2,1,0}(u_3) \\
& + G_{0,v_2,-1+v_2}(u_3) - G_{v_2,-1+v_2,-1+v_2}(u_3)
      + G_{-v_1+v_2,-1+v_2,-1+v_2}(u_3) \Bigr]
- G_{0,-1,0}(u_3) \\
&
+ G_{-v_1,-1,0}(u_3) + G_{-v_1,1-v_1,0}(u_3) + G_{-v_1,1-v_1,1-v_1}(u_3) - G_{1-v_1,1-v_1,0}(u_3) \\
&
- G_{1-v_1,1-v_1,1-v_1}(u_3) - G_{v_2,v_2,0}(u_3) 
- G_{v_2,v_2,-1+v_2}(u_3) + G_{0,-1,-1+v_2}(u_3) \\
&
+ G_{v_2,-1+v_2,0}(u_3) + G_{v_2,0,-1+v_2}(u_3)
- G_{0,-v_1+v_2,1-v_1}(u_3) - G_{0,-v_1+v_2,-1+v_2}(u_3) \\
&
- G_{-v_1,-1,-1+v_2}(u_3) - G_{-v_1,-v_1+v_2,1-v_1}(u_3)
- G_{-v_1,-v_1+v_2,-1+v_2}(u_3) - G_{1-v_1,v_2,0}(u_3) \\
&
- G_{1-v_1,v_2,-1+v_2}(u_3) + G_{1-v_1,-v_1+v_2,1-v_1}(u_3)
+ G_{1-v_1,-v_1+v_2,-1+v_2}(u_3) + G_{v_2,0,1-v_1}(u_3)\\
&
- G_{v_2,1-v_1,0}(u_3) - G_{v_2,1-v_1,-1+v_2}(u_3)
- G_{v_2,-1+v_2,1-v_1}(u_3) + G_{v_2,-v_1+v_2,1-v_1}(u_3) \\
&
+ G_{v_2,-v_1+v_2,-1+v_2}(u_3) - G_{-v_1+v_2,0,1-v_1}(u_3)
- G_{-v_1+v_2,1-v_1,0}(u_3) - G_{-v_1+v_2,0,-1+v_2}(u_3) \\
&
- G_{-v_1+v_2,-1+v_2,0}(u_3) + G_{-v_1+v_2,1-v_1,-1+v_2}(u_3) +  G_{-v_1+v_2,-1+v_2,1-v_1}(u_3) \\
&
- G_{0,(1-v_1)v_2/(1-v_1+v_1 v_2),0}(u_3) + G_{v_2,(1-v_1)v_2/(1-v_1+v_1 v_2),0}(u_3) \\
&
+ G_{0,(1-v_1)v_2/(1-v_1+v_1 v_2),1-v_1}(u_3) - G_{v_2,(1-v_1)v_2/(1-v_1+v_1 v_2),1-v_1}(u_3) \,.
\end{aligned}
\label{D1explicit}
\ee
The weight 4 function $D_0$ is more complicated, but we 
provide $\RR_4^{(2)}(u_1;u_3,v_1,v_2)$ in terms
of $G$-functions in the ancillary file {\tt R42\_rational.m}.
The basis of lower-weight functions appearing in the coproducts
of $\RR_4^{(2)}$ is given on the rational surface
in the ancillary file {\tt Prat.m}.


\subsubsection{FFOPE limit: Near multi-collinear factorization}

\label{subsubsec:factorization}

In the multi-collinear limit, scattering amplitudes and form factors
are expected to factorize into splitting amplitudes and lower-multiplicity
amplitudes and/or form factors (see e.g. ref.~\cite{Bern:2008ap}).
For the four-point form factors the kinematical limits can be expressed in
terms of the OPE variables as $T \to 0$ and/or $T_2 \to 0$.
The limit $T \to 0$ corresponds to the triple collinear limit where
$p_4, p_1, p_2$ are all parallel ($p_4 \parallel p_1 \parallel p_2$),
which sends $v_4,u_4,u_1 \to 0$.
The limit $T_2 \to 0$ corresponds to the ordinary collinear limit
$p_1 \parallel p_2$, which sends $u_1 \to 0$~\cite{Dixon:2022xqh}.

The triple collinear limit, where the form factor is related to the MHV
six-gluon amplitude~\eqref{eq:6pt_tricol},
has already been used to give boundary
information for the remainder function on the
rational surface.
The ordinary collinear limit, like the soft limit,
reduces the four-point form factor down to a three-point
form factor~\eqref{eq:3ptff}. Moreover, it has recently been shown,
at symbol level, that these multi-collinear limits relate an antipodal
self-duality for $\RR_4$ to the duality between the MHV six-point amplitude
and the three-point form factor~\cite{Dixon:2022xqh}.

We now consider the double series expansion in both $T$ and $T_2$.
The power series in $T_2$ at leading $T^0$ order corresponds to the
OPE expansion of the six-gluon amplitude found in the triple-collinear limit.
The power series of $T$ at leading $T_2^0$ order corresponds to the
FFOPE expansion of the three-point form factor found in the ordinary
collinear limit. Additional information is provided by
the terms with positive powers of both $T$ and $T_2$.
They are predicted by the four-point [FF]OPE
framework~\cite{Alday:2010ku,Basso:2013vsa,Basso:2013aha,Basso:2014koa,
Basso:2014jfa,Basso:2014nra,Belitsky:2014sla,Belitsky:2014lta,
Basso:2014hfa,Belitsky:2015efa,Basso:2015rta,Basso:2015uxa,
Belitsky:2016vyq,Basso:2023bwv}, more specifically
refs.~\cite{Sever:2020jjx,Sever:2021nsq,Sever:2021xga,Wilhelmprivate}.

In the OPE limit, the symbol alphabet,
written in terms of the OPE variables, simplifies to
\be
 \Phi_{\rm OPE} = \big \{
   S, \quad S_2,  \quad S^2+1,\quad S_2^2+1,\quad
   S^2 S^2_2 + S^2 +S_2^2,\quad S^4 S_2^2+S^4+2 S^2 S_2^2+S_2^2 \big\}.
\label{SS2alphabet}
\ee
There are also the infinitesimal letters $T$ and $T_2$, which just
generate powers of the logarithms $\ln T$ and $\ln T_2$.
To make the alphabet~(\ref{SS2alphabet})
\textit{linearly reducible} \cite{Duhr:2019tlz},
we find it convenient to use the variables
\be
 X = \frac{S^2}{1+S^2} , \quad Y = \frac{S_2^2}{1+S_2^2},
\ee
so that the alphabet becomes 
\be
 \Phi_{XY} = \big \{ X,\quad Y, \quad 1-X, \quad 1-Y, \quad X+Y-XY,
 \quad X^2+(1-X^2)Y \big \},
\ee
which is linearly reducible in the $\{Y,X\}$ basis.

We note that although $\Phi_{XY}$ appears to contain $X^2$,
after integrating in $Y$, the quadratic dependence on $X$ drops out.
More specifically, we follow the fibration algorithm described
in refs.~\cite{Duhr:2019tlz, Anastasiou:2013srw}.
After integration in $Y$, the alphabet in the next iteration is
\be
\Phi^{(Y)}_{XY} \equiv
\{ Q R' - R Q' | (QY+R) \in \Phi_{XY} , (Q'Y+R') \in \Phi_{XY}   \}
= \{ X, 1-X, 1+X \} \,,
\label{alphabetinX}
\ee
and this is linear in $X$. For instance, take the two letters
$X+Y-XY$ and $X^2+(1-X^2)Y$.  They contain $X^2$ but only
contribute to $\Phi^{(Y)}_{XY}$ as $ (1-X)X^2 - X (1-X^2) = -X(1-X)$.

Exploiting the linearity,
we are able to integrate up around the OPE limit
and fix the function-level information for the basis $P$
functions for the coproducts of $\RR_4^{(2)}$, iteratively in the weight,
by performing similar procedures as we did for the rational surface limit.

However, for higher powers of $T$ and $T_2$ in the OPE limit, we need
to fix more constants in the coproducts than were needed for the rational
surface.  In order to fix them, we used a combination of dihedral
symmetry constraints (see Sec.~\ref{sec:dihsymconstraints}) and
matching to the FFOPE data.
The ancillary file {\tt PTT2to0\_XY.m} contains the $G$-function
representation of these basis functions in the OPE limit, at leading power
in $T$ and $T_2$.  It contains the constants needed for taking derivatives
in the full five-dimensional phase space.

We have also constructed all these functions at higher orders in $T$ and
$T_2$, through order $T^2 T^2_2$.  At weight 4, we obtain the OPE limit
of $\RR_4^{(2)}$ itself.
The OPE expansion of the remainder function for terms with positive powers
of both $T$ and $T_2$, through order $T^2 T^2_2$, has the form:
\be
\begin{aligned}
 \RR_4^{(2)} =
 T^2   \bigg\{ & T_2   ( F_2 + F_2^{-1} )
      \Bigl(  \ln T \,  A_{1,1} + \ln T_2 \,  A_{1,2}
         + A_{1,3} + {\zeta_2}   A_{1,4} \Bigr) \\
& + T_2^2   \Bigl[ ( F_2^2 + F_2^{-2} )
  \Bigl( \ln T \, A_{2,2,1} +  \ln T_2 \, A_{2,2,2}
      + A_{2,2,3} + {\zeta_2} A_{2,2,4} \Bigr)  \\
& \quad \quad \quad \quad \quad
   + \ln T \, A_{2,0,1} + \ln T_2 \, A_{2,0,2}
   + A_{2,0,3} + {\zeta_2} A_{2,0,4} \Bigr] \bigg\}
\ +\ {\cal O}(T^2 T_2^3) \,.
\end{aligned}
\label{R42_FFOPE_terms}
\ee
The $A$ coefficients depend only on $S,S_2$ (or $x=S^2,y=S_2^2$).
They are given in Appendix~\ref{app:nff}.  We expose the beyond-the-symbol
terms containing $\zeta_2$ explicitly as $A_{1,4}$, $A_{2,2,4}$, $A_{2,0,4}$.
To compare with the FFOPE results~\cite{Wilhelmprivate}, which are provided
as a series expansion around $S\to 0$, $1/S_2\to 0$, we also perform this
straightforward series expansion.  The available FFOPE components
are $A_{1,i}$, $i=1,2,3,4$; $A_{2,2,1}$, $A_{2,2,2}$, and $A_{2,0,2}$.
They all agreed perfectly with the series expansions of our results.

\subsubsection{2D kinematics}

When the external momenta are constrained to lie in two spacetime dimensions,
the corresponding periodic Wilson loop has been computed at strong
coupling~\cite{Maldacena:2010kp}.  Here we consider the weak-coupling value.
In terms of Mandelstam variables, this limit is parametrized by
\be
\begin{aligned}
    u_1 &= v_1(1-v_2),
    \quad  u_2 = v_1 v_2 ,
    \quad u_3 = (1-v_1) v_2,  \quad u_4 = (1-v_1)(1-v_2), \\
    \quad v_3 &= 1-v_1, \quad v_4 = 1-v_2 \,.
\end{aligned}
\label{2Dkin}
\ee
The symbol alphabet in this limit simplifies to
\be
\Phi_{\rm 2D} =
\big \{ v_1,\ v_2,\ 1-v_1,\ 1-v_2,\ 1-v_1-v_2,\ v_1 - v_2\big\}.
\label{2Dalphabet}
\ee
(In principle, the alphabet could also contain the letters
$1+v_1-v_2$ and $1-v_1+v_2$, but they cancel in the symbol of the
remainder function.)

The 2D kinematics make contact with the rational surface
(and its dihedral image)
on the four
one-parameter line segments $v_1 = 0$, $v_1 = 1$, $v_2=0$, and $v_2=1$.
These segments form a square boundary of the region $0 \leq v_1,v_2 \leq 1$.
One of the segments, $v_2 = 1$, is a one-parameter subspace of the
$p_1 \to 0$ soft limit, where the behavior of $\RR_4^{(2)}$ is specified by
a limit of \eqn{eq:3ptff}.  Another segment, $v_1 = 0$, is a subspace
of the triple-collinear limit, where the behavior of the remainder function
is dictated by a limit of \eqn{eq:6pt_tricol}. 

Furthermore, this square parametrized by $(v_1,v_2)$ is mapped into
itself by the bulk cyclic transformation $\mathcal{C}$.
Projected onto this surface, the transformation is
\bea
&&\mathcal{C}: \ \ v_1 \to v_2 \to ( v_3 = 1-v_1)
   \to (v_4 = 1-v_2) \to v_1, \label{C2D}\\
&&\mathcal{C}^2: \ \ v_1 \lr 1-v_1, \quad v_2 \lr 1-v_2.  \label{Csq2D}
\eea
We require this dihedral symmetry of the remainder function to hold
at function level for 2D kinematics. 

Based on these constraints, we can fully recover the function-level
information in 2D kinematics. The remainder function in
this limit can be expressed in terms of $G$-functions of $v_1,v_2$.
The detailed expression is given in Appendix~\ref{app:2dkin}.


\subsection{Dihedral symmetry constraints}
\label{sec:dihsymconstraints}

We have made heavy use of dihedral symmetry in the evaluation of the
remainder function in the aforementioned boundary kinematics.
We have further used the information from the cycle/flip symmetry that
connects different boundaries, in order to fix $\RR_4^{(2)}$ in the bulk.
That is, we provide constants for all the lower-weight coproducts
required to integrate up $\RR_4^{(2)}$ off the rational surface at function
level.  In the OPE parametrization, near the OPE limit,
the constants are provided via
the ancillary file~{\tt PTT2to0\_XY.m}.  In the cross-ratio parametrization
on the rational surface, the constants are provided via
the ancillary file~{\tt Prat.m}.

The rest of this subsection collects how various dihedral symmetries act in
relevant limits.

In the OPE limit, there is a one-parameter line parametrized by $S$
in the limit that $S_2=\frac{1}{T_2} \to \infty$.
The flip transformation $\mathcal{F}$ in \eqn{eq:flip}
is represented on this line by
\be 
{\cal F}:\ \ S \to \frac{1}{S}. 
\ee

As mentioned above, there is a two-parameter subsurface~(\ref{ratu3to0})
of the rational surface, which is mapped to itself by the
reflection~(\ref{eq_cyc_flip}).  It is also mapped to itself by the
$\mathcal{C}^2$ symmetry (the cycle symmetry \eqref{eq:cyc} applied twice).
In this limit, $\mathcal{C}^2$ acts as 
\be
u_1 \lr u_3, \quad v_1 \lr 1-v_2 \,.
\ee
Compare this with the $\mathcal{C}^2$ symmetry in 2D kinematics, \eqn{Csq2D}.
They are compatible because the intersection has either $v_1,v_2 \to 0$
or $v_1,v_2 \to 1$, so $v_1 \approx v_2$ in both cases.

The rational surface and the OPE limit make contact at one kinematic point,
namely $(u_1,u_2,u_3,v_1,v_2) = (0,1,0,1,1)$.
In the OPE parametrization, however,
this point is actually a line in the limit $T,T_2,S \to 0$,
which is parametrized by $S_2$.
In terms of the cross ratios $u_i,v_i$, the variable $S_2$
parametrizes how fast $u_3 \to 0$ compared to $v_2\to 1$.
The relation between the OPE variables and $u_i,v_i$ in this limit is
\be
u_1 \to T^2 T^2_2,\quad u_3 \to S^2,
\quad v_1 \to 1- T^2, \quad v_2 \to 1- \frac{S^2}{S_2^2} \,,
\ee
where $S_2$ is finite and $T,T_2,S$ are small.  Hence $u_3/(1-v_2) \to S_2^2$.

The cycle-then-flip symmetry~(\ref{eq_cyc_flip}), which maps the
$u_3\to 0$ rational surface to the $u_1 \to 0$ rational surface,
is also preserved near this kinematic point (or the line parametrized by $S_2$).

\section{Remainder function in the bulk}
\label{sec:bulk}

In this section, we give numerical results on several slices through
the rational surface. We also check the proposed antipodal
self-duality~\cite{Dixon:2022xqh} at the function level,
by comparing the $\zeta_3$ parts of the derivatives (or $\{3,1\}$ coproducts)
and the branch cuts (to obtain the antipodally related $\{1,3\}$ coproducts).

\subsection{Slices through the rational surface}

\begin{figure}
    \centering
    \includegraphics[width=7cm]{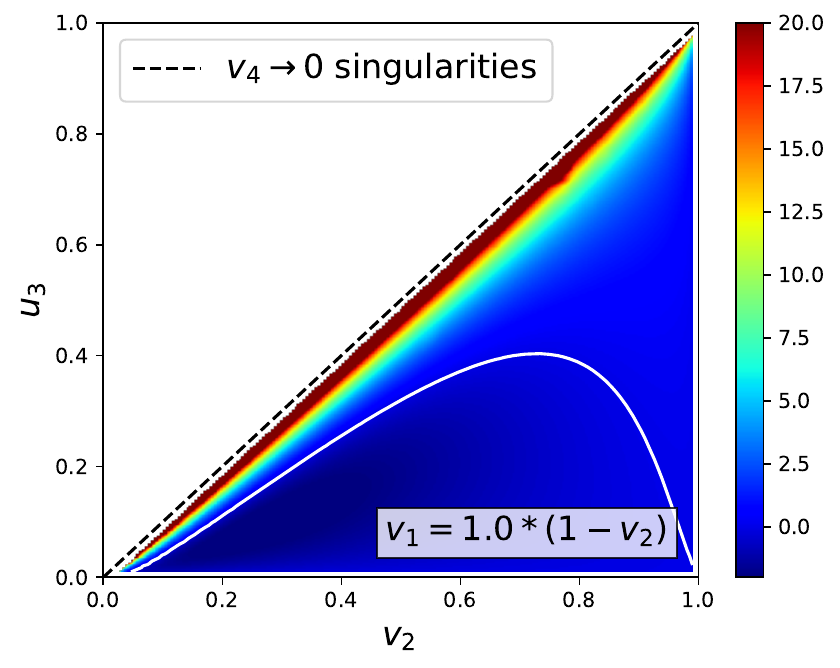}
    \includegraphics[width=7cm]{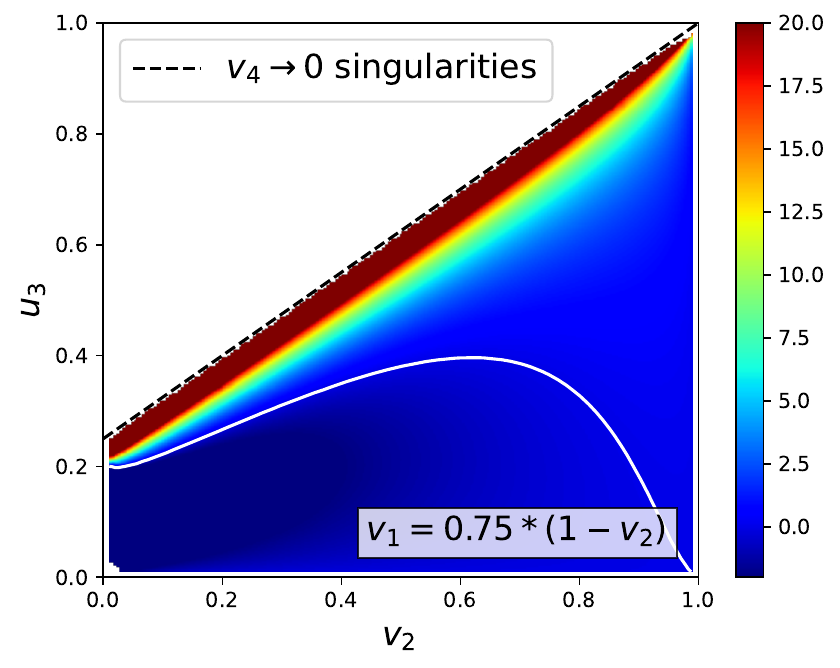}
    \includegraphics[width=7cm]{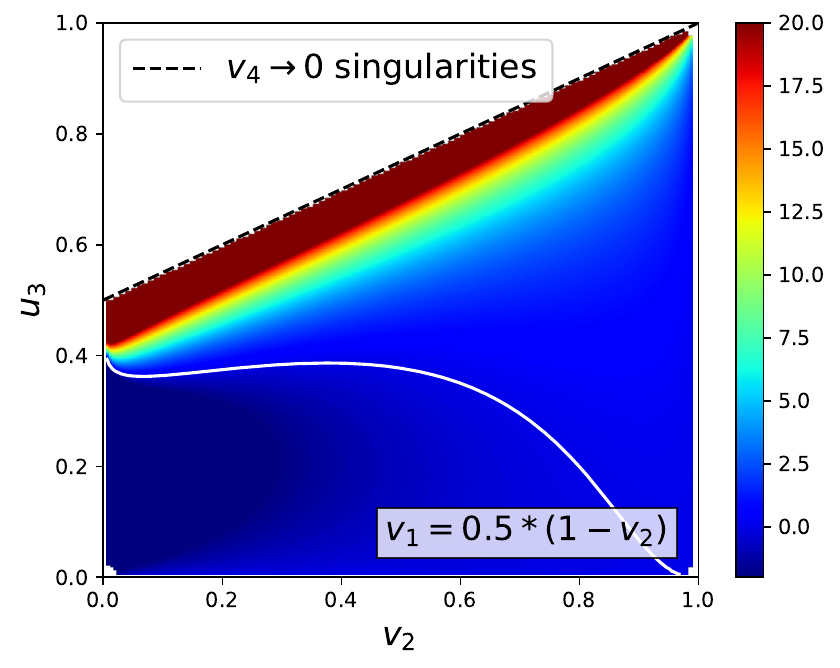}
    \includegraphics[width=7cm]{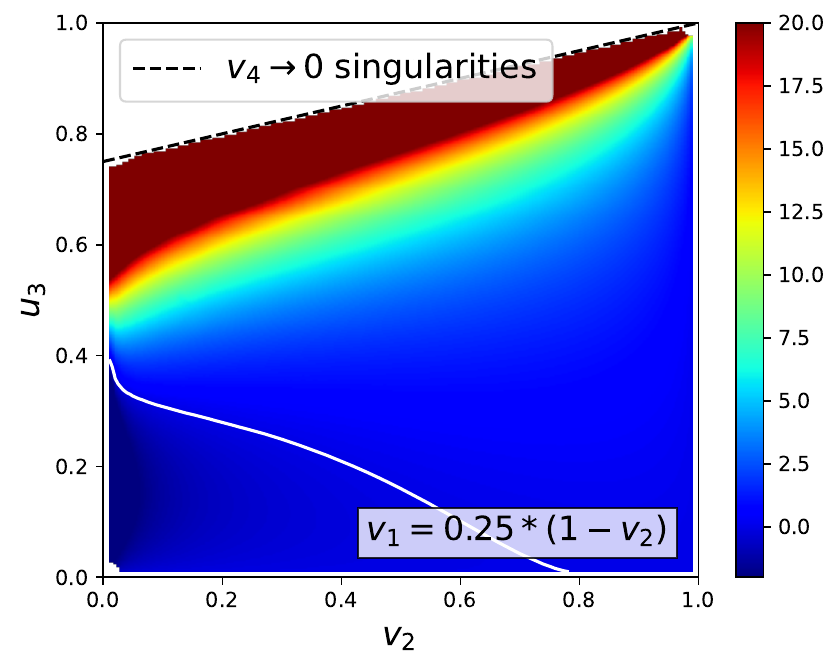}
    \caption{The finite part $D_0(u_3,v_1,v_2)$ of the remainder function
    $\RR_4^{(2)}$ on the ${v_1} = k({1-v_2})$ slice with the
    constant $k=$ 0.25, 0.5, 0.75, 1.
    The white curves highlight where $\RR_4^{(2)}$ flips sign.}
    \label{fig:triplecolside}
\end{figure}

\begin{figure}
    \centering
    \includegraphics[width=7cm]{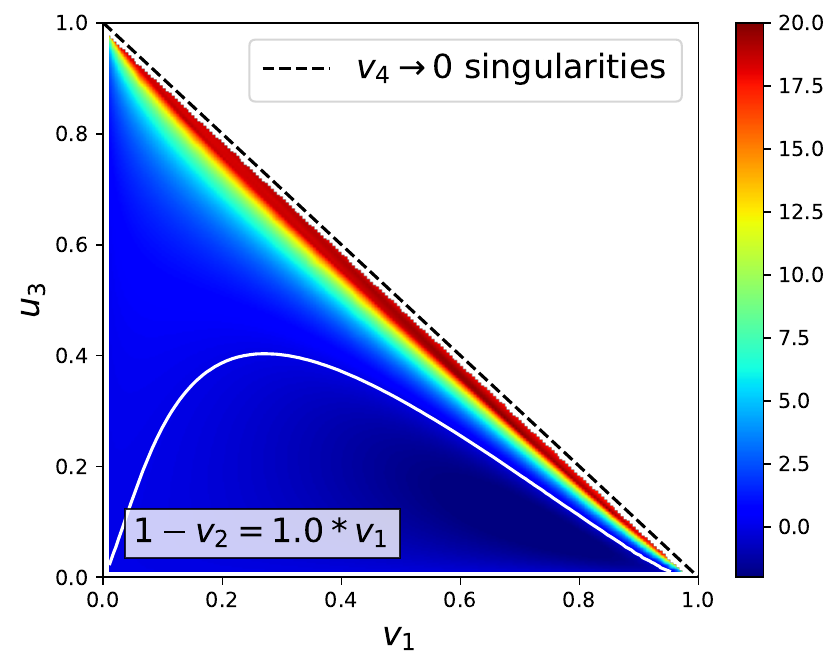}
    \includegraphics[width=7cm]{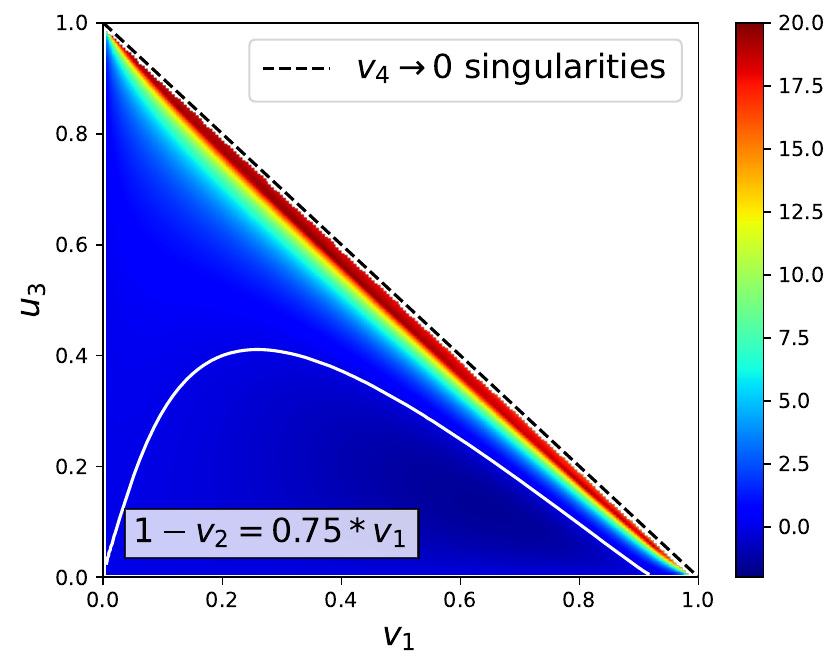}
    \includegraphics[width=7cm]{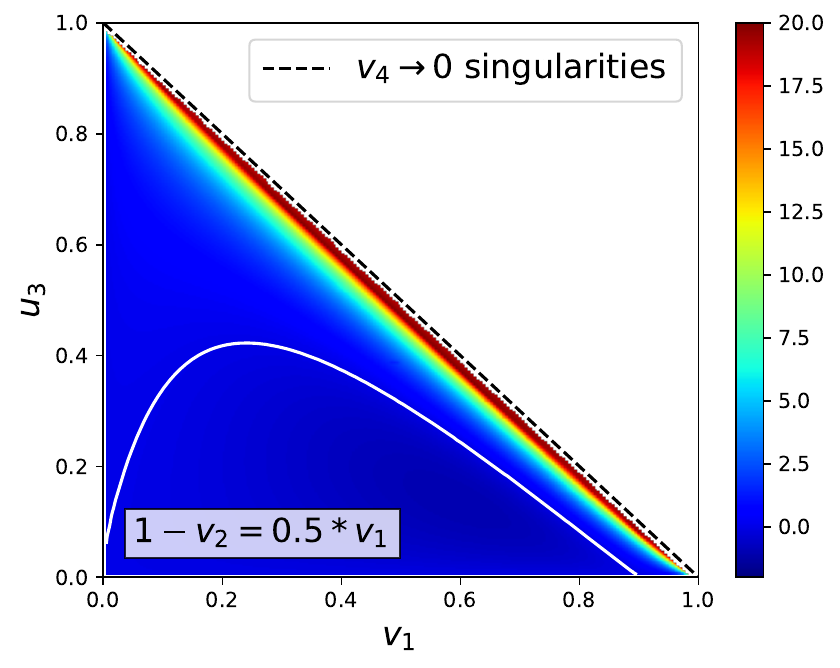}
    \includegraphics[width=7cm]{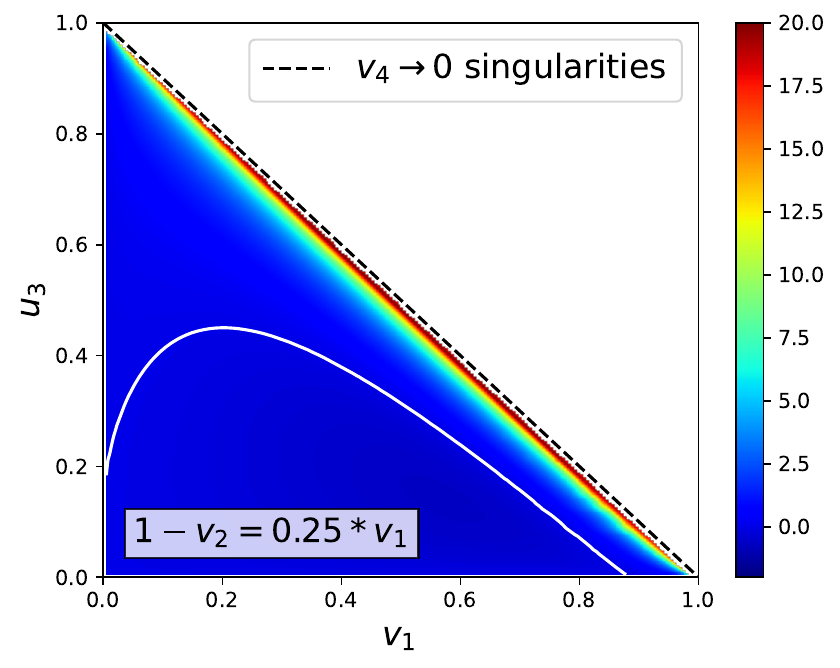}
    \caption{The finite part of the remainder function $\RR_4^{(2)}$
    on the ${v_1} = k({1-v_2})$ slice when $k \ge 1$. }
    \label{fig:softside}
\end{figure}

We present numerical values of $\RR_4^{(2)}$ in the three-parameter
rational surface kinematics introduced in Sec.~\ref{subsubsec:rational} by
evaluating them on several two-parameter slices through this surface.
As mentioned in Sec.~\ref{sec:uplift}, $\RR_4^{(2)}$ has a mild (linear)
logarithmic singularity in $u_1$ on the rational surface,
$\RR_4^{(2)}|_{\rm rat.} = D_0 + D_1 \, \ln u_1 $, as given in \eqn{D0D1decomposition}.
In Figs.~\ref{fig:triplecolside} and \ref{fig:softside},
we plot the finite part $D_0(u_3,v_1,v_2)$ as a function of $u_3$ and $v_2$,
on slices parametrized by $v_1 = k(1-v_2)$,
for different values of the constant $k$.
We see that the remainder function is smooth
inside this region and diverges only near a physical singularity, indicated
by the dashed line at $u_3 = 1 - k(1-v_2)$,
where the variable $v_4 \to 0$. 

\subsection{Antipodal duality beyond the symbol}

Antipodal duality between the three-point form factor
and the MHV six-gluon amplitude has been verified well beyond the level
of the symbol~\cite{Dixon:2021tdw}.  Because the right-hand side of
the coaction is only defined modulo $i\pi$ (and any powers thereof), at present
all the checks have been modulo $i\pi$.  This still leaves a large number
of beyond-the-symbol checks that have been passed.
For example, at dual points where the
three-point form factor and the MHV six-gluon amplitude evaluate to
MZVs, in the $f$-alphabet
representation~\cite{Brown:2011ik,HyperlogProcedures}
of the MZV's, the values of the multi-loop form factor and amplitude
are related to each other by dropping all $\pi$-containing terms
and reversing the order of all the $f$ letters, through eight
loops~\cite{Dixon:2021tdw,Dixon:2023kop}.

On the other hand, the antipodal self-duality of the four-point form factor
has only been checked so far at the symbol level~\cite{Dixon:2022xqh},
since the function is only now available.  We want to show invariance
under the antipode map, reversing the symbol and the various coactions,
combined with the \emph{kinematic map}:
\begin{equation}
T \to \sqrt{\frac{T_2}{S_2}} \,, \qquad S \to \sqrt{\frac{1}{T_2S_2}} \,,
\qquad T_2 \to \frac{T}{S} \,, \qquad S_2 \to \frac{1}{TS} \,,
\label{eq:kinmap}
\end{equation}
Because antipodal duality holds between the three-point form factor
and the MHV six-gluon amplitude including multiple zeta values, and because these
two functions are limits of the four-point form factor, it would be surprising
if antipodal-self-duality failed for $\RR_4^{(2)}$ at the function level.
Still, we should check it.

At two loops, or weight four, there is not too much that can be checked
beyond the symbol level, given the modulo $i\pi$ constraint.
The only constant at or below weight four, that does not vanish modulo $i\pi$,
is $\zeta_3$. There are two ways that $\zeta_3$ can appear in the coaction:
\begin{enumerate}
\item In the $\{3,1\}$ part of the coaction, one can go to a particular point where
the weight three function evaluates to $\zeta_3$, and look for terms
of the form $c \, \zeta_3 \otimes \ln\phi'$, where $\phi'$ is a symbol letter
(that appears as a final entry).
\item In the $\{1,3\}$ part of the coaction, one can look for terms
of the form $c \, \ln\phi \otimes \zeta_3$, where $\phi$ is a symbol letter
(that appears as an initial entry).
\end{enumerate}
To perform the function level check at two loops, we simply need to show
that these two types of terms map into each other under the antipode, which
exchanges $\{3,1\} \leftrightarrow \{1,3\}$, and under the kinematic map, which
should exchange letters $\phi \leftrightarrow \phi'$, while keeping the constant
prefactors $c$ the same.

In fact we will show that this duality of $\zeta_3$ terms holds by
mapping vanishing constant prefactors to each other, $c=0$.
This happens at the \textit{OPE limit} point where
$(u_1,u_2,u_3,v_1,v_2) = (0,1,0,1,1)$.
This point is a fixed point of the kinematic map~(\ref{eq:kinmap}), when we
take $T,T_2 \to 0$, and $S\to0$, $S_2\to\infty$.  To see this, we can
let $T \propto T_2^2$, $S \propto T_2$, $S_2 \propto 1/T_2^3$.
Then both sides of the map~(\ref{eq:kinmap}) scale the same way as $T_2 \to 0$.

At the OPE limit point, as discussed in Sec.~\ref{subsubsec:factorization},
$\RR_4^{(2)}$ vanishes at leading power in $T,T_2$.
{In addition, inspection of all of the $\{3,1\}$ coproducts at the OPE limit point, using the ancillary file {\tt PTT2to0\_XY.m}, shows that none of them contains a $\zeta_3$.  Such a $\zeta_3$ controls derivatives around the OPE limit point, and so it would also have shown up in the OPE expansion~\eqref{R42_FFOPE_terms_app}.} Thus there is no $\zeta_3 \ln(\phi')$ term for any letter $\phi'$. Also, at the OPE limit point $u_1,u_3,u_4,v_3,v_4$ all vanish, and so the logarithms of these five letters have branch cuts originating there.
Therefore we also know that the $\{1,3\}$ coproducts $\ln\phi \otimes \zeta_3$
have vanishing coefficient for $\phi\in\{u_1,u_3,u_4,v_3,v_4\}$;
otherwise we would see a logarithmic singularity in $\RR_4^{(2)}$ at
the OPE limit point, directly on the edge of the Euclidean sheet.

The task of this subsection is to show that the constants
in $c \, \ln\phi \otimes \zeta_3$ for the other
three first entries, $\phi \in\{u_2,v_1,v_2\}$, also vanish.
Because $u_2,v_1,v_2 \to 1$ at the OPE limit point, the logarithms
vanish there. In order to
reveal these discontinuities, we need to connect the OPE limit point with
limits where $u_2,v_1$ or $v_2$ vanishes.

\begin{figure}
    \centering
    \includegraphics[width=0.4\textwidth]{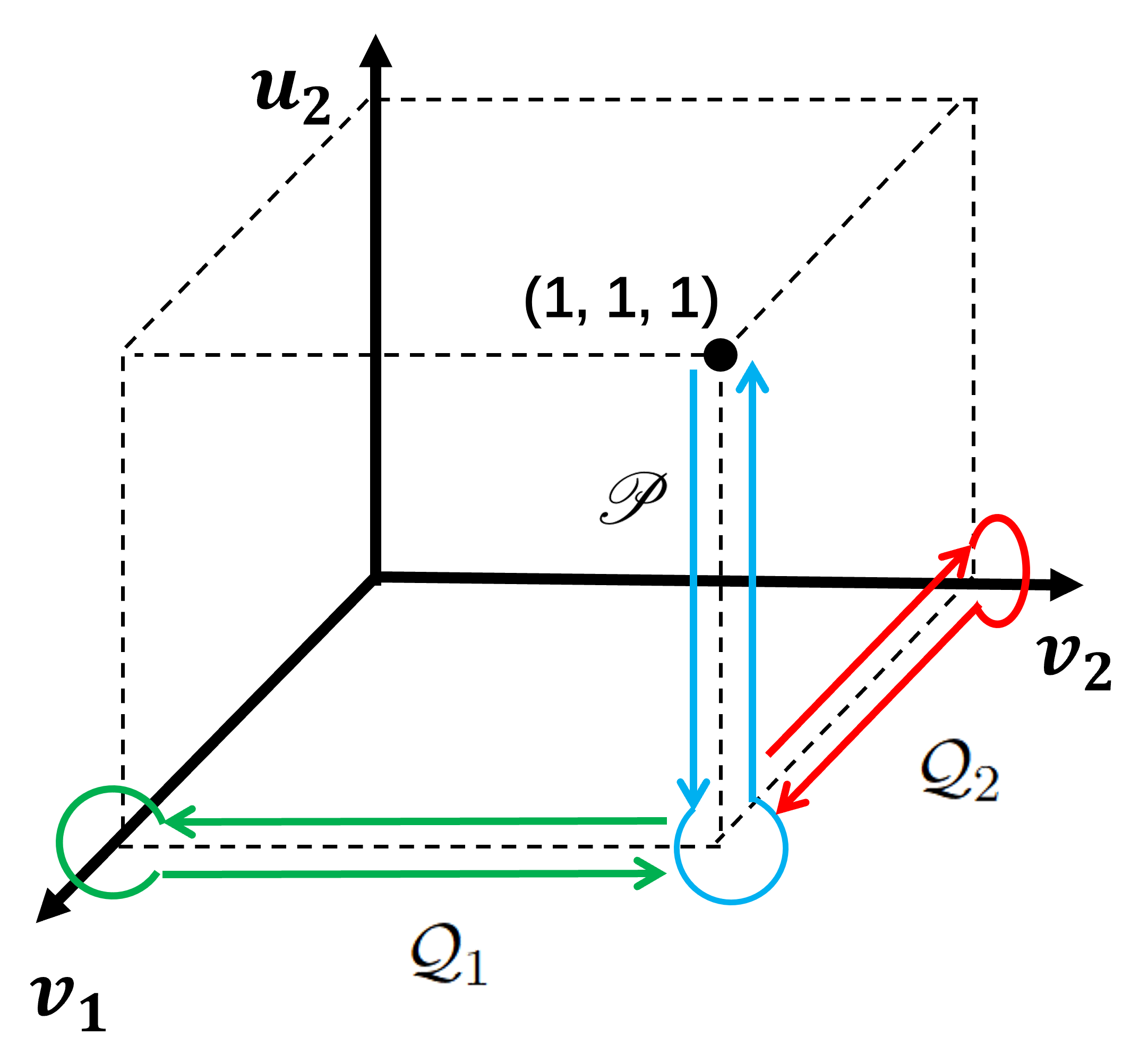}
    \caption{Diagram showing paths along which we integrate
    to obtain the $u_2,v_1,v_2$ discontinuities of $\RR_4^{(2)}$.}
    \label{fig:disc}
\end{figure}

First we consider $\phi = u_2$.  We reveal the potential
$\ln u_2 \otimes \zeta_3$ term by considering a one-dimensional path in the
phase-space which is parametrized by $u_2$ and which
extends out of the rational surface:
\be
\mathscr{P}:\ u_1,u_3 \to 0,\quad v_1,v_2 \to 1,\quad 0<u_2<1.
\ee 
Along this path $\mathscr{P}$, shown in Fig.~\ref{fig:disc},
the symbol alphabet simplifies to
$\{ \eta, 1+ \eta, 1- \eta \}$ with $\eta = \sqrt{1-u_2}$,
i.e.~the $u_2$ dependence becomes HPLs of $\eta$ with
indices $a_i\in\{0,1,-1\}$.
The following logarithms are divergent on this path $\mathscr{P}$:
\{$\ln u_1$, $\ln u_3$, $\ln(1-v_1)$, $\ln(1-v_2)$\}.

Any term with one of these divergent logarithms cannot contribute to the
$\ln u_2 \otimes \zeta_3$ term, because it would have to have a total weight
of at least five, and $\RR_4^{(2)}$ only has weight four.
Thus we only need to keep track of the finite part, which we compute to be:
\be
\begin{aligned}
\RR^{(2)}_4 &|_{u_1,u_3 \to 0, v_1,v_2 \to 1, \, {\rm finite\, part}} \\
=& -4 H_{-1, -1, 0, -1} (\eta) + 4 H_{-1, -1, 0, 0} (\eta)
+  4 H_{-1, -1, 0, 1} (\eta) + 4 H_{-1, -1, 1, 0} (\eta) \\
&  +  12 H_{-1, 0, 0, -1} (\eta) - 8 H_{-1, 0, 0, 0} (\eta)
-  12 H_{-1, 0, 0, 1} (\eta) + 4 H_{-1, 1, -1, 0} (\eta) \\
&  +  4 H_{-1, 1, 0, -1} (\eta) - 4 H_{-1, 1, 0, 0} (\eta)
-  4 H_{-1, 1, 0, 1} (\eta) - 4 H_{0, -1, -1, 0} (\eta) \\
&  +  8 H_{0, -1, 0, -1} (\eta) - 8 H_{0, -1, 0, 1} (\eta)
-  4 H_{0, -1, 1, 0} (\eta) + 8 H_{0, 0, -1, 0} (\eta) \\
&  -  24 H_{0, 0, 0, -1} (\eta) + 24 H_{0, 0, 0, 1} (\eta)
-  8 H_{0, 0, 1, 0} (\eta) - 4 H_{0, 1, -1, 0} (\eta) \\
&  -  8 H_{0, 1, 0, -1} (\eta) + 8 H_{0, 1, 0, 1} (\eta)
-  4 H_{0, 1, 1, 0} (\eta) + 4 H_{1, -1, 0, -1} (\eta) \\
&  -  4 H_{1, -1, 0, 0} (\eta) - 4 H_{1, -1, 0, 1} (\eta)
-  4 H_{1, -1, 1, 0} (\eta) - 12 H_{1, 0, 0, -1} (\eta) \\
&  +  8 H_{1, 0, 0, 0} (\eta) + 12 H_{1, 0, 0, 1} (\eta)
-  4 H_{1, 1, -1, 0} (\eta) - 4 H_{1, 1, 0, -1} (\eta) \\
&  +  4 H_{1, 1, 0, 0} (\eta) + 4 H_{1, 1, 0, 1} (\eta).
\end{aligned}
\label{R_pathP}
\ee
To compute the discontinuity in $u_2$ at $u_2=0$, we use the fact that
$u_2 \approx 2(1-\eta)$ in this limit.  We take the discontinuity
in~\eqn{R_pathP} at $\eta=1$ by clipping the index ``1'' off the back of
the HPL index list, and obtain:
\be
\begin{aligned}
{\rm disc}_{u_2}\Bigl[ \RR^{(2)}_4(\eta)|_{\rm finite\, part} \Bigr] &=
\pm i\pi \Bigl[ 4 H_{-1, -1, 0} (\eta) - 12 H_{-1, 0, 0} (\eta)
- 4 H_{-1, 1, 0} (\eta) - 8 H_{0, -1, 0} (\eta) \\
& + 24 H_{0, 0, 0} (\eta)
+ 8 H_{0, 1, 0} (\eta) - 4 H_{1, -1, 0} (\eta) + 12 H_{1, 0, 0} (\eta)
+ 4 H_{1, 1, 0} (\eta) \Bigr] \,.
\end{aligned}
\label{disc_R_pathP}
\ee
As we return to the OPE limit point, $\eta \to 0$, the discontinuity
behaves like,
\be
{\rm disc}_{u_2}\Bigl[ \RR^{(2)}_4(\eta)|_{\rm finite\, part} \Bigr] \to 
\pm 4 \pi i \ln^3 \eta \,,
\label{disc_R_pathP_eta0}
\ee
i.e.~$\zeta_3$ does not appear.
Hence there is no $\ln u_2 \otimes \zeta_3$ term
in the $\{1,3\}$ coproduct of $\RR^{(2)}_4$ at this point.

To reveal possible terms of the form
$\ln v_1 \otimes \zeta_3$ and $\ln v_2 \otimes \zeta_3$,
we need to take the two-segment paths $\mathscr{P}\mathscr{Q}_i$
shown in Fig.~\ref{fig:disc}.
From the OPE limit point $(u_1,u_2,u_3,v_1,v_2) = (0,1,0,1,1)$,
we first use the same path $\mathscr{P}$, where $u_2$ varies,
to reach the limit $(u_1,u_2,u_3,v_1,v_2) = (0,0,0,1,1)$.
This portion is shown in blue in Fig.~\ref{fig:disc}.
Then we vary either $v_1$ or $v_2$ to connect to either
$(u_1,u_2,u_3,v_1,v_2) = (0,0,0,1,0)$ or
$(u_1,u_2,u_3,v_1,v_2) = (0,0,0,0,1)$
via the segments $\mathscr{Q}_1$
(green) or $\mathscr{Q}_2$ red, respectively, in Fig.~\ref{fig:disc}.
The $v_1$ and $v_2$ discontinuities appear at the endpoints of these segments.

The symbol alphabet on the $u_1,u_2,u_3 \to 0$, $0<v_1,v_2<1$ surface,
which contains the green and red segments, simplifies to
$\{ v_1, v_2, 1-v_1, 1-v_2, 1-v_1-v_2, v_1+v_2 \}$.
The finite part of the remainder function can be represented on this surface
in terms of $G$-functions, as shown in Appendix~\ref{app:u0}.
Near $v_2\to 1$, the expression becomes simple logarithms,
\be
\begin{aligned}
\RR^{(2)}_{4,{\rm finite}}|_{u_i \to 0, v_2 \to 1} =
&\  \frac 18 \ln^4 (1-v_1) - \frac 12 \ln^3(1-v_1) \ln v_1
+ \frac 34 \ln^2(1-v_1) \ln^2 v_1 \\
& - \frac 12 \ln(1-v_1) \ln^3v_1 + \frac 18 \ln^4 v_1
-  \frac 14 \ln^2(1-v_1) \ln^2 (1-v_2) \\
& + \frac 12 \ln(1-v_1) \ln v_1 \ln^2(1-v_2)
- \frac 14 \ln^2 v_1 \ln^2(1-v_2) +\frac 18 \ln^4(1-v_2).
\end{aligned}
\label{alongQ1}
\ee
Then, when we cycle around $v_1 = 0$,
by letting $\ln v_1 \to \ln v_1 \pm i\pi$, and return to the point
$(u_2,v_1,v_2)=(0,1,1)$
along the red segment $\mathscr{Q}_2$, no $\zeta_3$ appears at $(0,1,1)$.
Furthermore, all such discontinuities at $(0,1,1)$ either vanish,
or else they contain the divergent logarithms $\ln(1-v_i)$.
Because of the overall weight constraint, the latter terms
cannot produce a $\zeta_3$ when transported along $\mathscr{P}$.
Hence no $\zeta_3$ appears at our base point $(u_2,v_1,v_2)=(1,1,1)$.
The $v_2$ discontinuity can be obtained similarly by using the $v_1 \lr v_2$
symmetry.  Therefore, the $\ln v_1 \otimes \zeta_3$ and
$\ln v_2 \otimes \zeta_3$ terms also vanish at the OPE limit point.
This concludes our check of antipodal self-duality in $\RR^{(2)}_4$
beyond the level of the symbol.

\section{Conclusion}
\label{sec:concl}

In this paper we determined the two-loop four-point form factor of the chiral
stress-energy tensor in planar $\mathcal{N}=4$ SYM~\cite{Dixon:2022xqh} in the
Euclidean region at function level.  We did so by giving the
iterated table of coproducts up through weight 4 and explicit $G$-function
representations of the basis functions in boundary kinematics.
We made use of the branch cut and dihedral symmetry constraints, as well as
several boundary kinematics where the behavior of the remainder function
is known.  In particular, we made use of a three-parameter rational surface
($u_1 \to 0, u_2 \to v_1 v_2$), which connects the soft limit,
triple collinear limit, 2D kinematics, and OPE limit.
The symbol alphabet is rationalized everywhere inside this rational surface.
The simplicity on this surface enables us to express the remainder function,
as well as the lower weight functions for bulk coproducts, as $G$-functions
with indices that are simple functions of the cross ratios; see for
example \eqn{D1explicit}.

By imposing the branch cut and symmetry constraints, and matching to known
boundary kinematics, we could fix all the zeta values on the rational surface.
A similar procedure in general bulk kinematics then gave us the entire table
of iterated coproducts at function level, including coproducts necessary for
taking derivatives in directions off of the rational surface.
The results are presented in the ancillary files. We further evaluated the
remainder function in several kinematic regions.  Plots illustrating the
numerical value of the remainder function on 2D slices of the rational surface
are presented in Sec.~\ref{sec:bulk}.  Also, by carrying the zeta-valued
information along various lines, to several other limits in the bulk,
we can get all the branch cut discontinuities of the remainder function.
Returning to a particular base point, the OPE limit point,
after taking such discontinuities, further checks the proposed antipodal
self-duality~\cite{Dixon:2022xqh} at function level.

With the full function for the two-loop four-point form factor,
several further investigations of kinematic regions can be performed.
Two such regions are multi-regge kinematics (MRK) and self-crossing kinematics.
In MRK, where there is a large rapidity separation between outgoing gluons,
the six-gluon remainder function exhibits a factorization after a
Fourier-Mellin transformation,
both for $2\to4$ kinematics~\cite{Lipatov:2010qf,Fadin:2011we}
and $3\to3$ kinematics~\cite{Bartels:2010tx}.  This representation has been
further exploited to give all-loop order resummations at
next-to-next-to-leading logarithmic
approximations~\cite{Dixon:2012yy, Pennington:2012zj},
and soon thereafter to give all subleading
logarithmic orders~\cite{Basso:2014pla}.
The four-gluon form factor has a MRK-like region in $2\to 3$ scattering,
where the operator is in the final state, and is emitted centrally between
two high-energy final state gluons.  
The phenomenon of gluon reggeization should be
universal~\cite{Caron-Huot:2013fea}, even in the presence of this operator,
which couples to two gluons. Therefore we expect a similar Fourier-Mellin
factorized form for the four-point form factor.  The details of such
a factorization remain to be studied in detail.

The Wilson loops dual to scattering amplitudes and form factors develop
singularities when the Wilson lines intersect.  The scattering kinematics
are similar to multi-parton scattering in hadronic collisions,
but here there are only two incoming gluons,
each of which splits into two almost on-shell collinear
virtual partons that then scatter off each other~\cite{Dixon:2016epj}.
(In the case of the four-point form factor,
one pair annihilates into an operator.)
For the case of six-gluon scattering, the self-crossing limit is
characterized by letting one cross ratio $\hat{u}\to 1$,
while the other two cross ratios are set equal to each other,
$\hat{w}=\hat{v}$.  The four-point form factor self-crossing kinematics
constrain three of the five kinematic variables.  They are a bit simpler
to describe in the OPE variables\footnote{We thank Benjamin Basso for
a discussion on this point.}:
$F_2 = 1$, and $T^2 = - S^2(1+T_2^2+S_2 T_2)/(S_2 T_2)$.
The logarithmic divergences as one approaches the self-crossing limit
can be characterized by a renormalization group
equation~\cite{Korchemsky:1993hr,Korchemskaya:1994qp}, 
$\mu \frac{\partial}{\partial \mu } \mathcal{W}_i
= -\Gamma^{ij} (\gamma, g) \mathcal{W}_j$,
where $\mathcal{W}_{1}$, $\mathcal{W}_{2}$ are the framed Wilson loops
with crossing and disconnected topology~\cite{Dixon:2016epj},
which mix with each other under renormalization. The renormalization scale
$\log \mu $ can be chosen as $\log(1-\hat{u})$ in the six-gluon case.
The cross anomalous dimension matrix $\Gamma^{ij}$ has kinematic
dependence only on the local crossing angle $\gamma$, and can be determined
by comparing with the six-gluon remainder function~\cite{Dixon:2016epj},
or using an all-orders formula based on
integrability~\cite{Caron-Huot:2019vjl}.
For the four-gluon form factor, we also expect a similar evolution
of self-crossing singularities, but again the details remain to
be investigated.

In conclusion, this paper serves to push forward the study of form factors
in planar $\mathcal{N} = 4$ SYM by providing function-level information
at two loops and four external legs, results which are complementary to
those in ref.~\cite{Guo:2024bsd}.  This information, and the methods used,
can be utilized further in future studies at higher multiplicity and
higher loop orders. 


\vskip0.5cm
\noindent {\large\bf Acknowledgments}
\vskip0.3cm

\noindent We thank Benjamin Basso, {\"O}mer G{\"u}rdo{\u{g}}an,
Zhengjie Li, Yu-Ting Liu, Andrew Mcleod, Matthias Wilhelm, and Gang Yang
for useful discussions. We thank Zhenjie Li and Song He for pointing out the error in Eq. (B.1).
We are grateful to the Simons Center for Geometry and Physics for hospitality
during the program ``Solving ${\cal N}=4$ super-Yang-Mills theory via
Scattering Amplitudes''.  LD also thanks Humboldt University Berlin
for hospitality.  This research was supported by the US Department of
Energy under contract DE--AC02--76SF00515.

\appendix

\section{Remainder function near the OPE limit}
\label{app:nff}

In this appendix, we provide the remainder function $\RR_4^{(2)}$
in the OPE limit, in a form which can be matched to the FFOPE data.
The cross ratios $u_i,v_i$ are expressed in terms of the OPE kinematic variables
$T,S,T_2,S_2,F_2$ using \eqn{eq:ope_parametrization}.
We perform a double series expansion in $T$ and $T_2$ around 0.
As mentioned in Sec.~\ref{subsubsec:factorization}, the terms at order
$T^0$ come from the OPE expansion of the MHV six-gluon amplitude and
are well-understood and checked to high loop orders.
Similarly, the terms at order $T_2^0$ come from the FFOPE expansion of the
three-point form factor, and are also well-understood.
Here we focus on terms with positive powers of both $T$ and $T_2$.
The first positive power of $T$ that contributes is $T^2$.
We provide this $T^2$ term, multiplied by either one or two powers of $T_2$.
At two loops, it has the following form,
repeated from \eqn{R42_FFOPE_terms} for convenience:
\be
\begin{aligned}
 \RR_4^{(2)} =
 T^2   \bigg\{ & T_2   ( F_2 + F_2^{-1} )
      \Bigl(  \ln T \,  A_{1,1} + \ln T_2 \,  A_{1,2}
         + A_{1,3} + {\zeta_2}   A_{1,4} \Bigr) \\
& + T_2^2   \Bigl[ ( F_2^2 + F_2^{-2} )
  \Bigl( \ln T \, A_{2,2,1} +  \ln T_2 \, A_{2,2,2}
      + A_{2,2,3} + {\zeta_2} A_{2,2,4} \Bigr)  \\
& \quad \quad \quad \quad \quad
   + \ln T \, A_{2,0,1} + \ln T_2 \, A_{2,0,2}
   + A_{2,0,3} + {\zeta_2} A_{2,0,4} \Bigr] \bigg\}
\ +\ {\cal O}(T^2 T_2^3) \,.
\end{aligned}
\label{R42_FFOPE_terms_app}
\ee
Here the coefficients $A_{1,i}$, $A_{2,2,i}$, and $A_{2,0,i}$, for $i=1,2,3,4$,
depend only on 
\be
    x = {S^2} , \quad y = {S_2^2} \,,
\ee
and contain both rational functions in $x,y$ and polylogarithms.
At two loops, all the polylogarithms are classical polylogarithms,
i.e.~${\rm Li}_n$ for $n=2,3$.  The expressions for $A_{1,3}$,
$A_{2,2,3}$ and $A_{2,0,3}$ are quite lengthy, so we provide them
instead in an ancillary file, {\tt R42\_OPE.txt}, which also
contains computer-readable forms of all the other coefficients.

The $A_{1,i}$ coefficients (except $A_{1,3}$) are given by
\be 
\begin{aligned}
    A_{1,1} = \frac{1}{\sqrt{y}} \Big[ &
     4   (x y+x+2 y+ \frac{y}{x})   \Big(
  {\rm Li}_{2} (\frac{x^2}{x^2 y+x^2+2 x y+y})
+ {\rm Li}_{2} (\frac{x}{(1+x)(x y+x+y)}) \\
& \quad \quad + {\rm Li}_{2} (\frac{-1}{x y+x+y})
  - {\rm Li}_{2} (\frac{x}{x y+x+y})
  + {\rm Li}_{2} (\frac{-x^2}{y(1+x)^2})
  - {\rm Li}_{2} (-\frac{1}{y}) \\
& \quad \quad + \ln y   \big( - 2 \ln (x^2 y+x^2+2 x y+y) + \ln (x y+x+y) + \ln (1+y)\\
     & \quad \quad + \ln x             + 2   \ln (1+x)\,\, \big)  + \frac 32   \ln^2 (x^2 y+x^2+2 x y+y) \\
     & \quad \quad - \ln (x^2 y+x^2+2 x y+y) \big( 3   \ln (1+x) + \ln (x y+x+y) + \ln x \big)    \\
     & \quad \quad  + \frac 12 \ln^2 (x y+x+y) + \ln (x y+x+y) \big( \ln (1+x) - \ln (1+y) \big)     + \frac 52 \ln^2 (1+x) \Big) \\
     & +\frac{4 y (1+x)^2}{x}   \ln (1+x)   \Big( 2   \ln x - 2   \ln (1+x) + 3 \Big) \\
     & - 8   (x y+x+2  y+\frac{y}{x})   \ln (x^2 y+x^2+2 x y+y)+ 12   x   \ln x + 4   x (y+1)   \ln (1+y) \\
     & + \frac{4 y (2 x y+2 x+y+2)}{x(y+1)}   \ln y + \frac{4 (x y+x+y)^2}{x(y+1)}   \ln (x y+x+y) \Big] \,,
\end{aligned}
\ee

\be
\begin{aligned}
    A_{1,2} = \frac{1}{\sqrt{y}} \Big[ &
    2   (x y+x+2 y+{\frac{y}{x}})   \Big(    {\rm Li}_{2} (\frac{x^2}{x^2 y+x^2+2 x y+y}) + {\rm Li}_{2} (\frac{-x^2}{y (1+x)^2})  \\
     & \quad \quad   +   \big(    \ln y + 2\ln (1+x) \big)   \big( {\frac{1}{2}}   \ln y + 2   \ln x + \ln (1+x)        - 2   \ln (x^2 y+x^2+2 x y+y) \big)   \\
     & \quad \quad  - 2   \ln x   \ln (x^2 y+x^2+2 x y+y)    + {\frac{3}{2}}   \ln^2 (x^2 y+x^2+2 x y+y) \Big) \\
    & + 2   (x y+x+2 y)   \big(    \ln y   \ln (1+y) -    \ln^2 (1+y) \big) \\
    & - 4   (x y+x+2 y+{\frac{y}{x}})   \ln (x^2 y+x^2+2 x y+y) + 8   x   \ln x  \\
    & + 8   \frac{y (1+x)^2}{x}   \ln (1+x) + 4 y (1+\frac{1}{x})   \ln y+ 4   (1+x) (1+y)   \ln (1+y)
    \Big] \,,
\end{aligned}
\ee


\be
\begin{aligned}
    A_{1,4} = \frac{1}{\sqrt{y}} \Big[ & 2 (2 x y + 2 x + 4 y + 3 \frac{y}{x} )
     \ln y+ 4 (3 x y+x+6 y+3 \frac{y}{x} )   \ln (1+x) \\
    & + 2 (x y+x+2 y) \ln (1+y)
    - 6  (x y+x+2 y+\frac{y}{x} )    \ln (x^2 y+x^2+2 x y+y)\\
    &- 4 + 4   x 
    \Big] \,.
\end{aligned}
\ee

The coefficients $A_{2,2,i}$ (except $A_{2,2,3}$) are given by
\be
\begin{aligned}
    A_{2,2,1} = &  - 4  \frac{(1+x)^2 y}{x}   \Big(    {\rm Li}_{2} (\frac{x^2}{x^2 y+x^2+2 x y+y}) + {\rm Li}_{2} (\frac{x}{(1+x)(x y+x+y)})  \\
    &\quad \quad + {\rm Li}_{2} (\frac{-1}{x y+x+y}) - {\rm Li}_{2} (\frac{x}{xy+x+y})  + {\rm Li}_{2} (\frac{-x^2}{y (1+x)^2}) - {\rm Li}_{2} (\frac{-1}{y})  \\
    &\quad \quad + \ln y   ( - 2   \ln (x^2 y+x^2+2 x y+y) + \ln (x y+x+y) + \ln (1+y) + \ln x              + 2   \ln (1+x) )\\
    &\quad \quad   + {\frac{3}{2}}   \ln^2 (x^2 y+x^2+2 x y+y)  - ( 3   \ln (1+x) + \ln (x y+x+y) + \ln x )   \ln (x^2 y+x^2+2 x y+y)\\
    &\quad \quad   + {\frac{1}{2}}   \ln^2 (x y+x+y) + ( \ln (1+x) - \ln (1+y) )   \ln (x y+x+y)  \\
    &\quad \quad + {\frac{1}{2}}   \ln (1+x)   ( \ln (1+x) + 4   \ln x ) \Big) \\
    & - 4   x   \ln x - 4   x y   \ln (1+y) - 8 \frac{y(1+x)^2}{x} \ln (1+x)- 4  \frac{ y (2 x y^2+2 x y+y^2+1+y)}{x (1+y)^2 }   \ln y \\
    & - 4   \frac{(x y+x+y)^2}{x(1+y)^2}   \ln (x y+x+y) \\
    & + 4    (x y+x+2 y+{\frac{y}{x}})    \ln (x^2 y+x^2+2 x y+y)
    + 4 \frac{y}{1+y} \,,
\end{aligned}
\ee

\be
\begin{aligned}
    A_{2,2,2} = & - 2   \frac{y (1+x)^2}{x}   \Big(    {\rm Li}_{2} (\frac{x^2}{x^2 y+x^2+2 x y+y}) + {\rm Li}_{2} (\frac{-x^2}{y (1+x)^2}) \\
    & \quad\quad + 2   ( {\frac{1}{2}}   \ln y + \ln (1+x) )   ( {\frac{1}{2}}   \ln y + 2   \ln x + \ln (1+x)                - 2   \ln (x^2 y+x^2+2 x y+y) )\\
    & \quad\quad   - 2   \ln x   \ln (x^2 y+x^2+2 x y+y) + {\frac{3}{2}}   \ln^2 (x^2 y+x^2+2 x y+y) \Big)\\
    &  - 2   y (2+x)   ( \ln y - \ln (1+y) )   \ln (1+y)
    - 2  \frac{(2 x^2 y-x^2+4 x y+2 y) x}{(1+x)^2 y}   \ln x \\
    &  - 6   \frac{y (1+x)^2}{x}   \ln (1+x)
    - \frac{(4 x y+3 y+3) y}{x (1+y)}   \ln y \\
    & - \frac{3 x y^3+5 x y^2+2 y^3+x y + 10 y^2-x+2 y+2}{y (1+y)}   \ln (1+y)\\
    & + \frac{(x^2 y+x^2+2 x y+y)(3 x^2 y-x^2+6 x y+3 y)}{(1+x)^2 x y}
       \ln (x^2 y+x^2+2 x y+y) + 2 \,,
\end{aligned}
\ee


\be
\begin{aligned}
    A_{2,2,4} = & -2 y (2 x+4 +\frac{3}{x}) \ln y
    - 12 \frac{y (1+x)^2}{x}   \ln (1+x)  \\
    &- 2 y (2+x)   \ln (1+y)+ 6 \frac{y (1+x)^2}{x}   \ln (x^2 y+x^2+2 x y+y) \\
    & - 4 x + \frac{3 x+2}{y(1+x)^2} + \frac{4 y}{y+1} \,.
\end{aligned}
\ee

Finally, the coefficients $A_{2,0,i}$ (except $A_{2,0,3}$) are given by
\be
\begin{aligned}
    A_{2,0,1} = & - 8  \frac{ (1+x)^2 y }{x}   \Big(   {\rm Li}_{2} (\frac{x^2}{x^2 y+x^2+2 x y+y}) + {\rm Li}_{2} (\frac{x}{(1+x)(x y+x+y)}) + {\rm Li}_{2} (\frac{-1}{x y+x+y}) \\
    & \quad \quad - {\rm Li}_{2} (\frac{-x}{y(1+x)})   - 2   {\rm Li}_{2} (\frac{x}{x y+x+y}) + {\rm Li}_{2} (\frac{-x^2}{y (1+x)^2}) - {\rm Li}_{2} (\frac{-1}{y})   + {\frac{3}{2}}   \ln^2 (x^2 y+x^2+2 x y+y)\\
    & \quad \quad - \ln^2 (x y+x+y) + \frac{3}{4}   \ln^2 (1+x)   - \ln y   \big( {\frac{1}{2}}   \ln y - 2   \ln (x y+x+y) \big)  \\
    & \quad \quad - ( \ln y - \ln (x y+x+y) )   ( \ln x + \ln (1+x) +\ln (1+y) ) + \frac{1}{4}   \ln (1+x)   ( \ln (1+x) + 4   \ln x )   \\
    & \quad \quad+ \ln y   \big( - 2   \ln (x^2 y+x^2+2 x y+y) + \ln (x y+x+y) + \ln (1+y) + \ln x               + 2   \ln (1+x) \big) \\
    & \quad \quad  - ( 3   \ln (1+x) + \ln (x y+x+y) + \ln x )   \ln (x^2 y+x^2+2 x y+y)   + \ln (\frac{1+x}{1+y} )   \ln (x y+x+y) \Big) \\
    & + 8   \frac{x}{y}   \Big(     {\rm Li}_{2} (\frac{x}{xy+x+y}) + {\rm Li}_{2} (\frac{-x}{y(1+x)})   + \big( \ln y - \ln (x y+x+y) \big)   \big( \ln x + \ln (1+x) + \ln (1+y) \big) \\
    & \quad \quad  + \ln y   \big( {\frac{1}{2}}   \ln y - 2   \ln (x y+x+y) \big)  + {\frac{1}{2}}   \ln^2 (1+x) + {\frac{3}{2}}   \ln^2 (x y+x+y) \Big) \\
    & + 16   (1+x)   \Big(     {\rm Li}_{2} (\frac{x}{xy+x+y}) + {\rm Li}_{2} (\frac{-x}{y(1+x)})   + {\frac{3}{2}}   \ln^2 (x y+x+y) - \frac{3}{4}   \ln^2 (1+x)   \\
    & \quad \quad + \ln y   \big( {\frac{1}{2}}   \ln y - 2   \ln (x y+x+y) \big) + \frac 14   \ln (1+x)   \big( \ln (1+x) + 4   \ln x \big)  \\
    & \quad \quad  + \big( \ln y - \ln (x y+x+y) \big)   \big( \ln x + \ln (1+x) + \ln (1+y) \big)  \Big) \\
    & + 8   (1+\frac{x}{y})   \ln x - 8   \frac{(y-1) (1+x)^2}{x}   \ln (1+x)  - 8   \frac{x y^3-x y^2-3 x y-y^2-x}{x (1+y)^2}   \ln y \\
    &  + 8   (2 x +y+1+\frac xy)   \ln (1+y)  - 8   \frac{(y^2+3 y+1) (x y+x+y)^2}{xy(1+y)^2}   \ln (x y+x+y) \\
    & + 8    (x y+x+2 y+{\frac{y}{x}})    \ln (x^2 y+x^2+2 x y+y)  + 8   \frac{y}{1+y} \,,
\end{aligned}
\ee

\be
\begin{aligned}
    A_{2,0,2} = & 4 \frac{  (x y+x+y)^2}{xy}   \Big(    {\rm Li}_{2} (\frac{-x}{x^2 y+x^2+2 x y+y}) + 2   {\rm Li}_{2} (\frac{x}{xy+x+y})  \\
    & \quad \quad - {\rm Li}_{2} (\frac{-1}{x y+x+y}) + {\rm Li}_{2} (\frac{-x}{y(1+x)}) + {\rm Li}_{2} (\frac{-1}{y})  \\
    & \quad \quad+ ( 2   \ln x - \ln (x y+x+y) - \ln (x^2 y+x^2+2 x y+y) )   \ln (1+x) \\
    & \quad \quad + \ln y   ( 2   \ln x + \ln y + \ln (1+x) + \ln (1+y) - 3   \ln (x y+x+y) ) \\
    & \quad \quad - \ln (x y+x+y)   ( 2   \ln x + \ln (x^2 y+x^2+2 x y+y) ) \\
    & \quad \quad + {\frac{1}{2}}   \ln^2 (x^2 y+x^2+2 x y+y) - \ln^2 (1+y) \\
    & \quad \quad + {\frac{1}{2}}   \ln^2 (1+x)  + {\frac{5}{2}}   \ln^2 (x y+x+y) \Big) \\
    & - 4  \frac{ (1+x)^2 y}{x}   \Big(    {\rm Li}_{2} (\frac{x^2}{x^2 y+x^2+2 x y+y})  + {\rm Li}_{2} (\frac{-x^2}{y (1+x)^2})  \\
    & \quad \quad +    (   \ln y + 2 \ln (1+x) )   ( {\frac{1}{2}}   \ln y + 2   \ln x + \ln (1+x)                        - 2   \ln (x^2 y+x^2+2 x y+y) )\\
    & \quad \quad  - 2   \ln x   \ln (x^2 y+x^2+2 x y+y) + \ln y   \ln (1+y) \\
    & \quad \quad - \ln^2 (1+y)  + {\frac{3}{2}}   \ln^2 (x^2 y+x^2+2 x y+y) \Big) \\
    & + 8  \frac{ x^3+x^2 y+x^2+2 x y+x+y} {y (1+x)^2}   \ln x- 8   \frac{y (1+x)^2}{x}   \ln (1+x) \\
    &  - 4   \frac{y^2-2 y-1}{1+y}   \ln y+ 4  \frac{ 2 x y^2+y^3+4 x y+2 x+3 y}{y(1+y)}   \ln (1+y) \\
    & - 8  \frac{ (x y+x+y)^2}{xy}   \ln (x y+x+y)\\
    & +  \frac{4(x^2 y+x^2+2 x y+y) (2 x^2 y+4 x y+x+2 y)}{(1+x)^2 x y}      \ln (x^2 y+x^2+2 x y+y)+ 8 \,,
\end{aligned}
\ee


\be
\begin{aligned}
    A_{2,0,4} = & - 4 \frac{3 x^2 y^2-6 x^2 y+6 x y^2-x^2-6 x y+3 y^2}{xy}   \ln (1+x) \\
    & - 12 \frac{(x y+x+y)^2}{xy}   \ln (x y+x+y) + 4 \frac{2 x y+x+2 y}{y}   ( 2 \ln y + \ln (1+y) ) \\
    & + 12 \frac{y (1+x)^2}{x}   \ln (x^2 y+x^2+2 x y+y)- 4 \frac{x^2}{y(1+x)^2} + 4 \frac{3 y+1}{y+1}  \,.
\end{aligned}
\ee

\section{Remainder function in 2D kinematics}
\label{app:2dkin}

When the external momenta all lie in the same 2D plane,
the remainder function reduces to $G$-functions of $v_1,v_2$:

\be
\begin{aligned}
    &\RR_4^{(2)} =  - 16 {\zeta_4}
+ 12 {\zeta_3} \Bigl[ G_{0} (v_1) + G_{1} (v_1) + G_{0} (v_2) +  G_{1} (v_2) \Bigr]\\
& + 4 {\zeta_2} \Bigl[ G_{0}^2 (v_1) + G_{0} (v_1) G_{0} (v_2)
     + G_{1} (v_1) G_{0} (v_2) + 2 G_{1} (v_1) G_{1} (v_2) + 4 G_{0} (v_1) G_{1} (v_2) + 3 G_{1} (v_2) G_{0} (v_2)\\
&\hskip1cm
     - 2 G_{0} (v_1) G_{0} (v_1-v_2) + G_{0}^2 (v_1-v_2)- 2 G_{0, 1} (1-v_2) - 2 G_{0, 1} \Bigl(\frac{-v_2}{v_1 - v_2}\Bigr)
    - 2 G_{-{v_1} + {v_2}, -1 + {v_2}} (v_2) \Bigr]\\
& + 2 G_{0}^3 (1-v_1) G_{0} (v_1) - 2 G_{1} (v_1) G_{0}^3 (v_1)
- 8 G_{1} (v_1) G_{0}^2 (v_1) G_{1} (v_2) - 4 G_{0}^2 (1-v_1) G_{0} (v_1) G_{1} (v_1+v_2)\\
& + 2 G_{1} (v_1) G_{0} (v_1) G_{1}^2 (v_1+v_2)
  + 4 G_{1} (v_1) G_{0}^2 (v_1) G_{0} (v_1-v_2)
  - 2 G_{1} (v_1) G_{0} (v_1) G_{0}^2 (v_1-v_2)\\
& + 2 G_{0}^2 (1-v_1) G_{0} (v_1) G_{0} (v_2)
  + 2 G_{1} (v_1) G_{0}^2 (v_1) G_{0} (v_2)
  - 8 G_{1} (v_1) G_{0} (v_1) G_{1} (v_2) G_{0} (v_2) \\
&- 4 G_{1} (v_1) G_{0} (v_1) G_{1} (v_1+v_2) G_{0} (v_2) + 
 2 G_{0} (v_1) G_{1}^2 (v_1+v_2) G_{0} (v_2) - 4 G_{1} (v_1) G_{0} (v_1) 
  G_{0} (v_1-v_2) G_{0} (v_2) \\
&+ 2 G_{1} (v_1) G_{0}^2 (v_1-v_2) G_{0} (v_2) - 
 4 G_{1} (v_1) G_{1} (v_2) G_{0}^2 (v_2) - 4 G_{0} (v_1) G_{1} (v_2) 
  G_{0}^2 (v_2) \\
& - 6 G_{1}^2 (v_2) G_{0}^2 (v_2) - 
 4 G_{0}^2 (1-v_1) G_{0, 1} (1-v_1) + 4 G_{0}^2 (v_1) G_{0, 1} (1-v_1) \\
& - 
 8 G_{1} (v_1) G_{1} (v_2) G_{0, 1} (1-v_1) + 
 8 G_{0} (v_1) G_{1} (v_2) G_{0, 1} (1-v_1) + 8 G_{1} (v_1) G_{1} (v_1+v_2) 
  G_{0, 1} (1-v_1) \\
&- 4 G_{1}^2 (v_1+v_2) G_{0, 1} (1-v_1) - 
 8 G_{0} (v_1) G_{0} (v_1-v_2) G_{0, 1} (1-v_1) + 
 4 G_{0}^2 (v_1-v_2) G_{0, 1} (1-v_1)  \\
 &- 
 4 G_{1} (v_1) G_{1} (v_2) G_{0, 1} (1-v_2) - 
 4 G_{0} (v_1) G_{1} (v_2) G_{0, 1} (1-v_2) + 4 G_{1} (v_1) G_{0} (v_2) 
  G_{0, 1} (1-v_2) \\
& + 4 G_{0} (v_1) G_{0} (v_2) G_{0, 1} (1-v_2) + 
 8 G_{1} (v_2) G_{0} (v_2) G_{0, 1} (1-v_2) - 8 G_{0, 1}^2 (1-v_2) \\
&+ 4 \Bigl[ G_{1} (v_1) \Bigl( G_{0} (v_1) - G_{0} (v_2) \Bigr)
     - 2 G_{0, 1} (1-v_1) \Bigr]
  G_{0, 1} \Bigl(\frac{-v_2}{v_1 - v_2}\Bigr)
     \\
&- 4 \Bigl[ G_{0} (v_1) \Bigl( G_{1} (v_1) + G_{0} (v_2) \Bigr)
     - 2 G_{0, 1} (1-v_1) \Bigr]
  G_{0, 1} \Bigl(\frac{-v_2}{1-v_1 - v_2}\Bigr)\\
&+ 4 G_{1} (v_1) \Bigl( G_{0} (v_1) - G_{0} (v_2) \Bigr)
     G_{-{v_1} + {v_2}, -1 + {v_2}} (v_2)
 - 8 G_{0, 1} (1-v_1) G_{-{v_1} + {v_2}, -1 + {v_2}} (v_2) \\
&- 4 G_{0} (v_1) \Bigl( G_{1} (v_1) + G_{0} (v_2) \Bigr)
      G_{-1 + {v_1} + {v_2}, -1 + {v_2}} (v_2)
 + 8 G_{0, 1} (1-v_1) G_{-1 + {v_1} + {v_2}, -1 + {v_2}} (v_2) \\
& + 24 G_{1} (v_2) \Bigl( G_{0, 0, 1} (1-v_1) + G_{0, 0, 1} (v_1) \Bigr)
  + 12 \Bigl( G_{0} (v_1) + G_{1} (v_1) + G_{0} (v_2) \Bigr) G_{0, 0, 1} (1-v_2) \\
& + 4 G_{1} (v_2) G_{0, 0, 1} (1-v_2)
  + 12 \Bigl( G_{0} (v_1) + G_{1} (v_1) + G_{1} (v_2) \Bigr) G_{0, 0, 1} (v_2)
  + 4 G_{0} (v_2) G_{0, 0, 1} (v_2) \\
&- 8 G_{0} (v_1) G_{0, 0, 1} \Bigl(\frac{v_2}{1 - v_1}\Bigr)
 - 8 G_{1} (v_1) G_{0, 0, 1} \Bigl(\frac{v_2}{v_1}\Bigr)
- 4 G_{1} (v_1)\Bigl( G_{-{v_1} + {v_2}, -1 + {v_2}, {v_2}} (v_2)
                     + G_{-{v_1} + {v_2}, {v_2}, -1 + {v_2}} (v_2) \Bigr)\\
& + 4 \Bigl( G_{0} (v_1) + G_{0} (v_2) \Bigr)
 \Bigl( G_{-1 + {v_2}, -{v_1} + {v_2}, {v_2}} (v_2)
     -  G_{-1 + {v_2}, -{v_1} + {v_2}, -1 + {v_2}} (v_2) \Bigr) \\
& + 4 \Bigl( G_{1} (v_1) + G_{0} (v_2) \Bigr)
   \Bigl( G_{-1 + {v_2}, -1 + {v_1} + {v_2}, {v_2}} (v_2)
        - G_{-1 + {v_2}, -1 + {v_1} + {v_2}, -1 + {v_2}} (v_2) \Bigr) \\
&- 4 G_{0} (v_1) G_{-1 + {v_1} + {v_2}, -1 + {v_2}, {v_2}} (v_2)
 - 4 G_{0} (v_1) G_{-1 + {v_1} + {v_2}, {v_2}, -1 + {v_2}} (v_2)
 - 16 G_{0, 0, 0, 1} (1-v_2) - 16 G_{0, 0, 0, 1} (v_2) \\
&- 8 G_{-1 + {v_2}, {v_2}, -{v_1} + {v_2}, -1 + {v_2}} (v_2) + 
 8 G_{-1 + {v_2}, {v_2}, -{v_1} + {v_2}, {v_2}} (v_2) - 
 8 G_{-1 + {v_2}, {v_2}, -1 + {v_1} + {v_2}, -1 + {v_2}} (v_2) \\
&+ 
 8 G_{-1 + {v_2}, {v_2}, -1 + {v_1} + {v_2}, {v_2}} (v_2) - 
 4 G_{-1 + {v_2}, -{v_1} + {v_2}, -1 + {v_2}, {v_2}} (v_2) - 
 4 G_{-1 + {v_2}, -{v_1} + {v_2}, {v_2}, -1 + {v_2}} (v_2) \\
&+ 
 8 G_{-1 + {v_2}, -{v_1} + {v_2}, {v_2}, {v_2}} (v_2) - 
 4 G_{-1 + {v_2}, -1 + {v_1} + {v_2}, -1 + {v_2}, {v_2}} (v_2) - 
 4 G_{-1 + {v_2}, -1 + {v_1} + {v_2}, {v_2}, -1 + {v_2}} (v_2) \\
&+ 
 8 G_{-1 + {v_2}, -1 + {v_1} + {v_2}, {v_2}, {v_2}} (v_2) \,.
\end{aligned}
\label{R42_2D}
\ee

The function is also available in Mathematica format in the ancillary file \verb|R42_2d.m|.

\section{Remainder function near $u_1,u_2,u_3 \to 0$}
\label{app:u0}

In order to connect the point $(u_1,u_2,u_3,v_1,v_2) = (0,0,0,1,1)$
to either $(0,0,0,0,1)$ or $(0,0,0,1,0)$, we integrated up the
remainder function
on the surface where $u_1,u_2,u_3 \to 0$, and $0<v_1,v_2<1$.
For convenience, we took the limit with $u_1 \ll u_3 \ll u_2$
(the precise hierarchy between the $u_i$ may not matter).
In this limit, the singular variable $u_2$ drops out, and
the remainder function has only a mild logarithmic divergence
involving $\ln u_1$ and $\ln u_3$: 
\be
{R}_{4}^{(2)}|_{u_i \to 0} = {R}_{4, \rm finite}^{(2)}
+ \ln u_1 {R}_{4, \rm div1}^{(2)} + \ln u_3 {R}_{4, \rm div 2}^{(2)}
+ \ln u_1 \ln u_3 {R}_{4, \rm div 3}^{(2)} \,.
\ee
The finite part is:
\be
\begin{aligned}
    {R}_{4, \rm finite}^{(2)} = & -G_{0, 0}(1-v_1) G_{0, 0}(1-v_2) + G_{0, 0}(1-v_2) G_{0, 1}(1-v_1)  +  G_{0, 0}(1-v_1) G_{0, 1}(1-v_2) \\
& - G_{0, 1}(1-v_1) G_{0, 1}(1-v_2)  +  G_{0, 0}(1-v_2) G_{1, 0}(1-v_1) - G_{0, 1}(1-v_2) G_{1, 0}(1-v_1) \\
& -  G_{0, 0}(1-v_2) G_{1, 1}(1-v_1) + G_{0, 1}(1-v_2) G_{1, 1}(1-v_1)  -  2 G_{0, 0}(1-v_1) G_{1, 1}(1-v_2) \\
& + 2 G_{0, 1}(1-v_1) G_{1, 1}(1-v_2)  +  2 G_{1, 0}(1-v_1) G_{1, 1}(1-v_2) - 2 G_{1, 1}(1-v_1) G_{1, 1}(1-v_2) \\
& +  2 G_{0, 0}(1-v_1) G_{1, v_1}(1-v_2) - 2 G_{0, 1}(1-v_1) G_{1, v_1}(1-v_2)  -  2 G_{1, 0}(1-v_1) G_{1, v_1}(1-v_2)\\
& + 2 G_{1, 1}(1-v_1) G_{1, v_1}(1-v_2)  +  G_{0, 0}(1-v_1) G_{v_1, 0}(1-v_2) - G_{0, 1}(1-v_1) G_{v_1, 0}(1-v_2) \\
& -  G_{1, 0}(1-v_1) G_{v_1, 0}(1-v_2) + G_{1, 1}(1-v_1) G_{v_1, 0}(1-v_2)  +  G_{0, 0}(1-v_1) G_{v_1, 1}(1-v_2) \\
&- G_{0, 1}(1-v_1) G_{v_1, 1}(1-v_2)  -  G_{1, 0}(1-v_1) G_{v_1, 1}(1-v_2) + G_{1, 1}(1-v_1) G_{v_1, 1}(1-v_2) \\
& -  2 G_{0, 0}(1-v_1) G_{v_1, v_1}(1-v_2) + 2 G_{0, 1}(1-v_1) G_{v_1, v_1}(1-v_2)  +  2 G_{1, 0}(1-v_1) G_{v_1, v_1}(1-v_2) \\
& - 2 G_{1, 1}(1-v_1) G_{v_1, v_1}(1-v_2)  -  G_{1}(1-v_2) G_{0, 0, 0}(1-v_1) + G_{v_1}(1-v_2) G_{0, 0, 0}(1-v_1) \\
& +  G_{1}(1-v_2) G_{0, 0, 1}(1-v_1) - G_{v_1}(1-v_2) G_{0, 0, 1}(1-v_1)  -  G_{0}(1-v_1) G_{0, 0, 1}(1-v_2) \\
& + G_{1}(1-v_1) G_{0, 0, 1}(1-v_2)  +  G_{0}(1-v_1) G_{0, 0, v_1}(1-v_2) - G_{1}(1-v_1) G_{0, 0, v_1}(1-v_2) \\
& +  G_{1}(1-v_2) G_{0, 1, 0}(1-v_1) - G_{v_1}(1-v_2) G_{0, 1, 0}(1-v_1)  -  G_{0}(1-v_1) G_{0, 1, 0}(1-v_2) \\
& + G_{1}(1-v_1) G_{0, 1, 0}(1-v_2)  -  G_{1}(1-v_2) G_{0, 1, 1}(1-v_1) + G_{v_1}(1-v_2) G_{0, 1, 1}(1-v_1) \\
& +  2 G_{0}(1-v_1) G_{0, 1, 1}(1-v_2) - 2 G_{1}(1-v_1) G_{0, 1, 1}(1-v_2)  -  G_{0}(1-v_1) G_{0, 1, v_1}(1-v_2) \\
& + G_{1}(1-v_1) G_{0, 1, v_1}(1-v_2)  +  G_{0}(1-v_1) G_{0, v_1, 0}(1-v_2) - G_{1}(1-v_1) G_{0, v_1, 0}(1-v_2) \\
& -  G_{0}(1-v_1) G_{0, v_1, 1}(1-v_2) + G_{1}(1-v_1) G_{0, v_1, 1}(1-v_2)  +  G_{1}(1-v_2) G_{1, 0, 0}(1-v_1) \\
& - G_{v_1}(1-v_2) G_{1, 0, 0}(1-v_1)  -  G_{0}(1-v_1) G_{1, 0, 0}(1-v_2) + G_{1}(1-v_1) G_{1, 0, 0}(1-v_2) \\
& -  G_{1}(1-v_2) G_{1, 0, 1}(1-v_1) + G_{v_1}(1-v_2) G_{1, 0, 1}(1-v_1)  +  G_{0}(1-v_1) G_{1, 0, 1}(1-v_2) \\
&- G_{1}(1-v_1) G_{1, 0, 1}(1-v_2)  -  G_{1}(1-v_2) G_{1, 1, 0}(1-v_1) + G_{v_1}(1-v_2) G_{1, 1, 0}(1-v_1) \\
& +  G_{1}(1-v_2) G_{1, 1, 1}(1-v_1) - G_{v_1}(1-v_2) G_{1, 1, 1}(1-v_1)  +  G_{0}(1-v_1) G_{1, v_1, 0}(1-v_2) \\
& - G_{1}(1-v_1) G_{1, v_1, 0}(1-v_2)  -  G_{0}(1-v_1) G_{1, v_1, 1}(1-v_2) + G_{1}(1-v_1) G_{1, v_1, 1}(1-v_2) \\
& +  G_{0}(1-v_1) G_{v_1, 0, 0}(1-v_2) - G_{1}(1-v_1) G_{v_1, 0, 0}(1-v_2)  -  G_{0}(1-v_1) G_{v_1, 0, v_1}(1-v_2) \\
& + G_{1}(1-v_1) G_{v_1, 0, v_1}(1-v_2)  +  G_{0}(1-v_1) G_{v_1, 1, 0}(1-v_2) - G_{1}(1-v_1) G_{v_1, 1, 0}(1-v_2) \\
& -  2 G_{0}(1-v_1) G_{v_1, 1, 1}(1-v_2) + 2 G_{1}(1-v_1) G_{v_1, 1, 1}(1-v_2)  +  G_{0}(1-v_1) G_{v_1, 1, v_1}(1-v_2) \\ \nonumber
\end{aligned}
\ee

\be
\begin{aligned}
& - G_{1}(1-v_1) G_{v_1, 1, v_1}(1-v_2)  -  2 G_{0}(1-v_1) G_{v_1, v_1, 0}(1-v_2) + 2 G_{1}(1-v_1) G_{v_1, v_1, 0}(1-v_2) \\
& +  2 G_{0}(1-v_1) G_{v_1, v_1, 1}(1-v_2) - 2 G_{1}(1-v_1) G_{v_1, v_1, 1}(1-v_2)  +  3 G_{0, 0, 0, 0}(1-v_1)\\
& + 3 G_{0, 0, 0, 0}(1-v_2)  -  3 G_{0, 0, 0, 1}(1-v_1) - G_{0, 0, 0, 1}(1-v_2) - 2 G_{0, 0, 0, v_1}(1-v_2) \\
& -  3 G_{0, 0, 1, 0}(1-v_1) - 2 G_{0, 0, 1, 0}(1-v_2)  +  3 G_{0, 0, 1, 1}(1-v_1) + 2 G_{0, 0, 1, v_1}(1-v_2) \\
& -  G_{0, 0, v_1, 0}(1-v_2) + G_{0, 0, v_1, 1}(1-v_2) - 3 G_{0, 1, 0, 0}(1-v_1) \\
& -  3 G_{0, 1, 0, 0}(1-v_2) + 3 G_{0, 1, 0, 1}(1-v_1) + G_{0, 1, 0, 1}(1-v_2) \\
& +  2 G_{0, 1, 0, v_1}(1-v_2) + 3 G_{0, 1, 1, 0}(1-v_1)  +  2 G_{0, 1, 1, 0}(1-v_2) - 3 G_{0, 1, 1, 1}(1-v_1) \\
& -  2 G_{0, 1, 1, v_1}(1-v_2) + G_{0, 1, v_1, 0}(1-v_2) - G_{0, 1, v_1, 1}(1-v_2) \\
& -  3 G_{1, 0, 0, 0}(1-v_1) - 4 G_{1, 0, 0, 0}(1-v_2)  +  3 G_{1, 0, 0, 1}(1-v_1) + 2 G_{1, 0, 0, 1}(1-v_2) \\
& +  2 G_{1, 0, 0, v_1}(1-v_2) + 3 G_{1, 0, 1, 0}(1-v_1)  +  2 G_{1, 0, 1, 0}(1-v_2) - 3 G_{1, 0, 1, 1}(1-v_1) \\
& -  2 G_{1, 0, 1, v_1}(1-v_2) + 2 G_{1, 0, v_1, 0}(1-v_2)  -  2 G_{1, 0, v_1, 1}(1-v_2) + 3 G_{1, 1, 0, 0}(1-v_1) \\
& +  2 G_{1, 1, 0, 0}(1-v_2) - 3 G_{1, 1, 0, 1}(1-v_1)  -  2 G_{1, 1, 0, 1}(1-v_2) - 3 G_{1, 1, 1, 0}(1-v_1) \\
& +  3 G_{1, 1, 1, 1}(1-v_1) - 2 G_{1, 1, v_1, 0}(1-v_2)  +  2 G_{1, 1, v_1, 1}(1-v_2) + 2 G_{1, v_1, 0, 0}(1-v_2) \\
& -  2 G_{1, v_1, 0, v_1}(1-v_2) - 2 G_{1, v_1, 1, 0}(1-v_2)  +  2 G_{1, v_1, 1, v_1}(1-v_2) + G_{v_1, 0, 0, 0}(1-v_2) \\
& -  G_{v_1, 0, 0, 1}(1-v_2) - G_{v_1, 0, v_1, 0}(1-v_2) + G_{v_1, 0, v_1, 1}(1-v_2) \\
& +  G_{v_1, 1, 0, 0}(1-v_2) + G_{v_1, 1, 0, 1}(1-v_2)  -  2 G_{v_1, 1, 0, v_1}(1-v_2) - 2 G_{v_1, 1, 1, 0}(1-v_2) \\
& +  2 G_{v_1, 1, 1, v_1}(1-v_2) + G_{v_1, 1, v_1, 0}(1-v_2)  -  G_{v_1, 1, v_1, 1}(1-v_2) - 2 G_{v_1, v_1, 0, 0}(1-v_2) \\
& +  2 G_{v_1, v_1, 0, v_1}(1-v_2) + 2 G_{v_1, v_1, 1, 0}(1-v_2)  -  2 G_{v_1, v_1, 1, v_1}(1-v_2) \,.
\end{aligned}
\ee

The divergent parts are
\be
\begin{aligned}
    {R}_{4, \rm div1}^{(2)} = &\ (G_{0} (1 - v_1 ) - G_{1} (1 - v_1 )) G_{0, 0} (1 - v_2 ) + (-G_{0} (1 - v_1 ) + G_{1} (1 - v_1 )) G_{0, 1} (1 - v_2 ) \\
     & + 
     G_{v_1 } (1 - v_2 ) (-G_{0, 0} (1 - v_1 ) + G_{0, 1} (1 - v_1 ) + G_{1, 0} (1 - v_1 ) - 
       G_{1, 1} (1 - v_1 )) \\
    & + G_{1} (1 - v_2 ) (G_{0, 0} (1 - v_1 ) - G_{0, 1} (1 - v_1 ) - 
       G_{1, 0} (1 - v_1 ) + G_{1, 1} (1 - v_1 )) \\
    & + (-G_{0} (1 - v_1 ) + G_{1} (1 - v_1 )) 
      G_{v_1 , 0} (1 - v_2 ) + (G_{0} (1 - v_1 ) - G_{1} (1 - v_1 )) G_{v_1 , 1} (1 - v_2 ) \\
    & - 
     G_{0, 0, 0} (1 - v_1 ) + 2 G_{0, 0, 0} (1 - v_2 ) + G_{0, 0, 1} (1 - v_1 ) - 
     G_{0, 0, 1} (1 - v_2 ) \\
    & - G_{0, 0, v_1 } (1 - v_2 ) + G_{0, 1, 0} (1 - v_1 ) - 
     G_{0, 1, 0} (1 - v_2 ) - G_{0, 1, 1} (1 - v_1 ) \\
    &+ G_{0, 1, v_1 } (1 - v_2 ) - 
     G_{0, v_1 , 0} (1 - v_2 ) + G_{0, v_1 , 1} (1 - v_2 ) + G_{1, 0, 0} (1 - v_1 ) \\
    &- 
     G_{1, 0, 0} (1 - v_2 ) - G_{1, 0, 1} (1 - v_1 ) + G_{1, 0, 1} (1 - v_2 ) - 
     G_{1, 1, 0} (1 - v_1 ) \\
    &+ G_{1, 1, 1} (1 - v_1 ) + G_{1, v_1 , 0} (1 - v_2 ) - 
     G_{1, v_1 , 1} (1 - v_2 ) - G_{v_1 , 0, 0} (1 - v_2 ) \\
    &+ G_{v_1 , 0, v_1 } (1 - v_2 ) + 
     G_{v_1 , 1, 0} (1 - v_2 ) - G_{v_1 , 1, v_1 } (1 - v_2 ) \,,
\end{aligned}
\ee
\be
\begin{aligned}
    {R}_{4, \rm div3}^{(2)} = &
    \ G_{0} (1 - v_2 ) (-G_{0} (1 - v_1 ) + G_{1} (1 - v_1 )) + 
     (G_{0} (1 - v_1 ) - G_{1} (1 - v_1 )) G_{1} (1 - v_2 )\\
     & - G_{0, 0} (1 - v_1 ) - 
     G_{0, 0} (1 - v_2 ) + G_{0, 1} (1 - v_1 ) + G_{0, v_1 } (1 - v_2 ) \\
     &+ G_{1, 0} (1 - v_1 ) + 
     G_{1, 0} (1 - v_2 )- G_{1, 1} (1 - v_1 ) - G_{1, v_1 } (1 - v_2 ) \,,
\end{aligned}
\ee
and ${R}_{4, \rm div2}^{(2)}$ is related to by ${R}_{4, \rm div2}^{(2)}$ by
the ``cycle-then-flip'' symmetry~(\ref{eq_cyc_flip}),
which exchanges $u_1 \lr u_3$, $v_1 \lr v_2$:
\be
    {R}_{4, \rm div2}^{(2)}(v_1,v_2) = {R}_{4, \rm div1}^{(2)}(v_2,v_1).
\ee
The finite part and the $\ln u_1 \ln u_3$ coefficient are both
invariant under $v_1 \lr v_2$.  Thus ${R}_{4}^{(2)}|_{u_i \to 0} $
preserves the cycle-then-flip symmetry~(\ref{eq_cyc_flip}).

\noindent

\bibliographystyle{JHEP}

\bibliography{R42}

\end{document}